\documentclass[twocolumn,nopacs,preprintnumbers,amsmath,amssymb,aps,eqsecnum]{revtex4}

\usepackage{graphicx}
\usepackage{dcolumn}
\usepackage{bm}
\usepackage[usenames,dvipsnames]{color}

\definecolor{altncolor}{rgb}{0,0,0.8}
\usepackage[colorlinks=true, linkcolor=green, anchorcolor=altncolor,
citecolor=altncolor, filecolor=altncolor, menucolor=altncolor,
urlcolor=altncolor]{hyperref}

\begin{document}


\author{Peter B. Weichman}

\affiliation{BAE Systems, Advanced Information Technologies, 600 District Avenue, Burlington, MA 01803}

\title{Competing turbulent cascades and eddy--wave interactions in shallow water equilibria}

\date{\today}

\begin{abstract}

In recent work, Renaud, Venaille, and Bouchet (RVB) \cite{RVB2016} revisit the equilibrium statistical mechanics theory of the shallow water equations, within a microcanonical approach, focusing on a more careful treatment of the energy partition between inertial gravity wave and eddy motions in the equilibrium state, and deriving joint probability distributions for the corresponding dynamical degrees of freedom. The authors derive a Liouville theorem that determines the underlying phase space statistical measure, but then, through some physical arguments, actually compute the equilibrium statistics using a measure that \emph{violates} this theorem. Here, using a more convenient, but essentially equivalent, grand canonical approach, the full statistical theory consistent with the Liouville theorem is derived. The results reveal several significant differences from the previous results: (1) The microscale wave motions lead to a strongly fluctuating thermodynamics, including long-ranged correlations, in contrast to the mean-field-like behavior found by RVB. The final effective model is equivalent to that of an elastic membrane with a nonlinear wave-renormalized surface tension. (2) Even when a mean field approximation is made, a rather more complex joint probability distribution is revealed. Alternative physical arguments fully support the consistency of the results. Of course, the true fluid final steady state relies on dissipative processes not included in the shallow water equations, such as wave breaking and viscous effects, but it is argued that the current theory provides a more mathematically consistent starting point for future work aimed at assessing their impacts.

\end{abstract}

\maketitle

\section{Introduction}
\label{sec:intro}

The modern era of exact statistical treatments of the late-time steady states of 2D fluid flows, properly accounting for the infinite number of conserved integrals of the motion, began with the Miller--Robert--Sommeria (MRS) theory of the 2D Euler equation \cite{M1990,RS1991,MWC1992,LB1967}, generalizing earlier approximate treatments going all the way back to the seminal work of Onsager \cite{O1949}, and progressing through the Kraichnan Energy--Enstrophy theory \cite{K1975}, and various formulations of the point vortex problem (see, e.g., \cite{MJ1974,LP1977}). Since then, the theory has been applied to significantly more complex systems, containing multiple interacting fields (in contrast to the Euler equation, which reduces to a single scalar equation for the vorticity), but still possessing an infinite number of conserved integrals \cite{HMRW1985}. These include, for example, magnetohydrodynamic equilibria \cite{JT1997,W2012}, 3D axisymmetric flow \cite{TDB2013}, and the shallow water equations \cite{WP2001,CS2002}, as well as numerous other geophysical applications \cite{BV2012}.

The theory of the shallow water system was recently revisited in Ref.\ \cite{RVB2016} (hereinafter referred to as RVB). The work highlighted simplifying approximations made in previous work on this system \cite{WP2001,CS2002}, and aimed to move beyond them in order to generate more quantitative predictions. Previous simplifications mainly involved the problem of dissipation of microscale gravity wave fluctuations. Such physical effects are certainly physically present, in the form of nonlinear phenomena such as wave breaking or shock wave dissipation, but lie beyond the shallow water approximation (which, in particular, assumes the length scale of horizontal motions to be much larger than the fluid depth). In previous work the small scale free surface fluctuations were simply set to zero at a convenient point in the calculation (citing untreated dissipation processes), and mean field variational equations describing the remaining large scale eddy motion were then derived \cite{WP2001}. In RVB, the shallow water system, though idealized, is taken at face value, and an attempt is made to treat the wave fluctuations in a more consistent manner, but also within a mean field approximation. The result is a very interesting equilibrium state that includes both steady large scale eddy motions and finite microscale wave fluctuations.

The key underlying physics here, also motivating earlier studies, is that the two nonlinearly interacting fields, surface height and eddy vorticity, when viewed in isolation, have very different turbulent dynamics. Two-dimensional eddy systems governed by Navier--Stokes turbulence tend to self-organize into long-lived, large-scale coherent structures such as cyclones (exemplified by Jupiter's Great Red Spot) and jets, a consequence of the famous 2D inverse energy cascade \cite{K1967,KM1980}. However, weak turbulence theory \cite{ZFL1998} predicts that interacting acoustic waves, similar to 3D Navier--Stokes turbulence, possesses a forward cascade of energy, transporting it from larger to smaller scales where it is ultimately acted upon by viscosity or other microscopic dissipation mechanisms. When both motions are present, the question arises as to what the final disposition of the energy is. The RVB results propose a quantitative answer, predicting the equilibrium distribution of energy (and other quantities of interest) between the large-scale eddy and microscale wave motions, depending of course on all of the conserved integral values set, for example, by a flow initial condition.

The purpose of the present paper is to revisit deeper simplifying mathematical assumptions made in RVB that strongly impact the derived statistical equilibrium state. Two key features are highlighted. First, the variational mean field results are at odds with other recent results for systems with multiple interacting fields which are only partially constrained by conservation laws (in contrast, e.g., to the Euler equation, in which the vorticity field completely specifies the dynamics, while at the same time its fluctuations are strongly limited by the conservation laws). For example, for magnetohydrodynamic equilibria, the unconstrained degrees possess finite microscale fluctuations that lead to a non-mean field thermodynamic description of the large scale flow \cite{W2012}. An analogous result is derived here: the surface height fluctuations are not controlled by the vorticity conservation laws, and lead to a strongly fluctuating equilibrium thermodynamics. Physically, the microscale surface height fluctuations lead to a fluctuating effective Coulomb-like interaction between vortices that does not self-average even on large length scales. A mean field description emerges only in an approximation where this effect is ignored.

Second, the formalism of statistical mechanics relies on identification of the correct phase space measure used to compute the thermodynamic free energy and perform statistical averages. This measure is determined by a Liouville theorem that characterizes the geometry of phase space flows. In particular, when expressed in terms of the correct combination of fields, these flows are incompressible, and this constrains the phase space measure to be a function only of the conserved integrals of the motion (expressed in terms of these particular field combinations). An issue addressed in this paper is that the correct Liouville theorem is indeed derived in RVB, but is not actually implemented correctly to define the phase space measure. The authors recognize this, but propose various physical arguments why their chosen implementation, which simplifies the mathematics (in particular, it makes the fields statistically independent), also makes more physical sense.

If the motivation of the study was to follow the full consequences of the shallow water equations, prior to speculating on the effects neglected physics, there appears to be a basic inconsistency here. In the following, the full statistical theory is derived using the correct equilibrium phase space measure. The resulting theory leads to much more complex behavior, and indeed has some unusual physical consequences---for example, the microscale fluctuations lead to an equilibrium-averaged flow that does \emph{not} satisfy the time-independent shallow water equations. Of course, which theory more closely reflects physical reality remains an interesting question, but the point of view taken here is that one should at least start by adhering as rigorously as possible to the mathematically consistent predictions of the model. Only following this should one attempt to insert physical considerations at various points to see what their affect might be. For example, a key consequence of the shallow water model is that the surface height fluctuations cascade to arbitrarily small wavelengths while at the same time maintaining a finite amplitude, thereby generating a kind of finite-thickness surface ``foam''. It is the dynamics of this foam that leads to both the strongly fluctuating equilibrium and to the violation of the time-independent equations, and was suppressed at the outset in previous work \cite{WP2001,CS2002}. These are consistent predictions of the model, but is obviously inconsistent with any physical final state, which must emerge by inserting a dissipation step to obtain a ``true'' equilibrium. How to best accomplish this lies beyond the scope of this paper, and would be an interesting topic for future work.

\subsection{Outline}
\label{sec:outline}

The aim of this paper is to formulate general statistical models of shallow water equilibrium states, and then explore some of their key, high-level features. More detailed, physically motivated, investigations of model predictions are left for future work.

The outline of the remainder of this paper is as follows. In Sec.\ \ref{sec:bkgnd} the shallow water equations are summarized, and the infinite number of conserved potential vorticity integrals are identified. In Sec.\ \ref{sec:canonfields} all quantities of interest are expressed in terms of the basic vorticity (velocity curl), compressional (velocity divergence), and fluid height fields. The free slip boundary conditions play a key role here, especially in multiply connected domains where a set of circulation integrals about each connected component of the boundary is separately conserved. The latter lead to an additional set of ``potential flow'' contributions to the energy, and also to the expressions for the linear or angular momentum (in the case of translation or rotation invariant domains, respectively, where they are conserved). These have not been previously considered in the context of the shallow water system. In Sec.\ \ref{sec:statmech} the equilibrium statistical mechanics formalism is introduced, with the conservation laws handled by introducing conjugate ``chemical potentials'' within the grand canonical approach. Application of a Kac--Hubbard--Stratanovich transformation allows one to exactly integrate out the fluid fields, and reduce the problem to that of a single effective field whose equilibrium average determines the large scale flow. The resulting statistical model is equivalent that of a fluctuating, scalar nonlinear elastic membrane problem \cite{W2012}. The model also has a dual description in terms of the vortex degrees of freedom interacting through a fluctuating Coulomb-like interaction. In Sec.\ \ref{sec:furtherprops}, we consider simplifying limits in which fluctuations are neglected. An approximate saddle point variational approach (analogous to, but quantitatively different from, that derived by RVB) is then used to illustrate further properties of the model. Equivalent forms of this theory are derived from both the elastic membrane and Coulomb models. The latter is closer in spirit to the RVB microcanonical approach. In Sec.\ \ref{sec:eulercomp} an interesting order-of-limits paradox (infinite gravity $g$ vs.\ perfect rigid lid Euler equation boundary condition) is examined. The two limits produce very different forms of the Liouville theorem, and the paradox is resolved in terms of the finite contribution of microscale gravity waves to the free energy due to the simultaneous divergence of the wave speed $c \approx \sqrt{g h}$. The paper is concluded in Sec.\ \ref{sec:conclude}. Two Appendices \ref{app:liouville} and \ref{app:liouvilleinequiv} prove a very general form of the Liouville theorem and review its relation to the statistical phase space integration measure. Some formal energy and momentum calculational details are relegated to App.\ \ref{app:KEPi}.

\begin{figure}
\includegraphics[width=3.0in,viewport=100 210 540 420,clip]{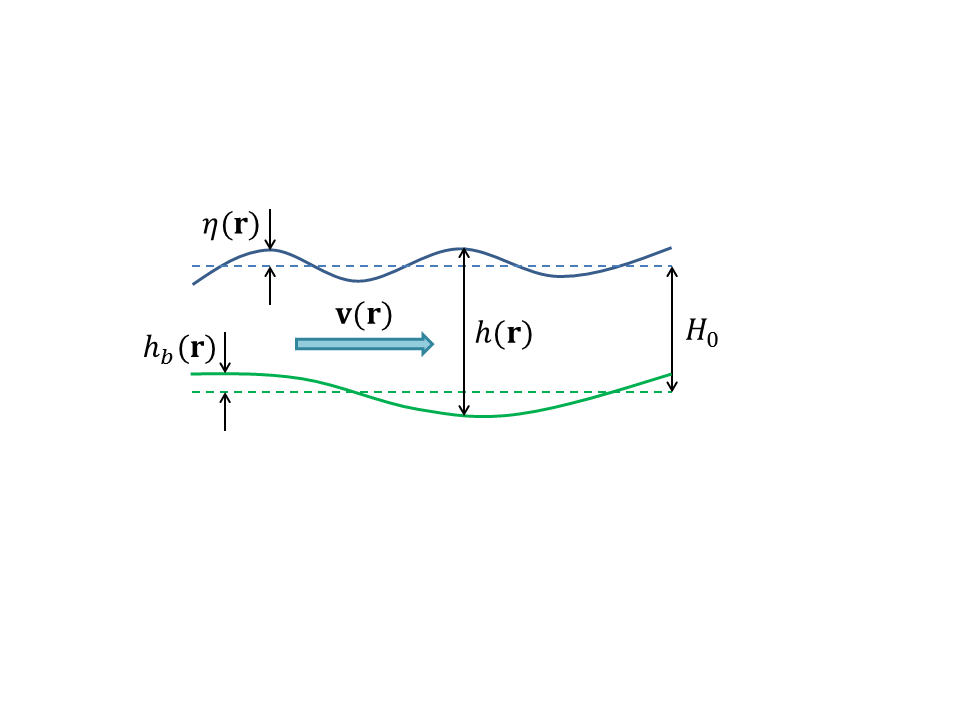}

\caption{Shallow water geometry and fields.}

\label{fig:swcartoon}
\end{figure}

\section{Background}
\label{sec:bkgnd}

The (2D) shallow water equations take the form \cite{foot:2Dcompress}
\begin{eqnarray}
\partial_t {\bf v} + ({\bf v} \cdot \nabla) {\bf v}
+ f {\bf \hat z} \times {\bf v} &=& -g \nabla \eta
\nonumber \\
\partial_t h + \nabla \cdot (h {\bf v}) &=& 0
\label{2.1}
\end{eqnarray}
where ${\bf v}$ is the (horizontal) velocity field, $h({\bf r})$ is the fluid layer thickness, $f({\bf r})$ is the Coriolis parameter, $h_b({\bf r})$ is the bottom height, and
\begin{equation}
\eta({\bf r}) = h({\bf r}) + h_b({\bf r}) - H_0
\label{2.2}
\end{equation}
is the surface height deviation from its average value
\begin{equation}
H_0 = \int_D \frac{d{\bf r}}{A_D} h({\bf r})
\label{2.3}
\end{equation}
(see Fig.\ \ref{fig:swcartoon}). Here $A_D$ is the area of the domain $D$, and we normalize the average bottom height to vanish,
\begin{equation}
\int_D \frac{d{\bf r}}{A_D} h_b({\bf r}) = 0.
\label{2.4}
\end{equation}
The second equation in (\ref{2.1}) expresses conservation of 3D fluid density through the mass current, or momentum (areal) density,
\begin{equation}
{\bf j} = \rho_0 h {\bf v}.
\label{2.5}
\end{equation}
The (fixed, uniform) fluid 3D mass density $\rho_0$ is included here for convenience in order to maintain a consistent set of physical units ($\rho_0$ drops out of the equations of motion).  One may simply set $\rho_0 = 1$ if one wishes.

\subsection{Conservation laws}
\label{sec:conslaws}

\subsubsection{Potential vorticity}
\label{subsec:potvort}

The potential vorticity,
\begin{equation}
\Omega = \frac{\omega + f}{h},\ \ \omega = \nabla \times {\bf v}
\label{2.6}
\end{equation}
which includes the combined effect of Earth and fluid rotation, is advectively conserved:
\begin{equation}
\frac{D\Omega}{Dt} \equiv \partial_t \Omega
+ ({\bf v} \cdot \nabla) \Omega = 0.
\label{2.7}
\end{equation}
It follows that, for any function $w(\Omega)$, $hw(\Omega)$ is a conserved density,
\begin{equation}
\partial_t [h w(\Omega)] + \nabla \cdot [h w(\Omega) {\bf v}] = 0,
\label{2.8}
\end{equation}
and hence that any integral of the form
\begin{equation}
I_w = \int_D d{\bf r} h({\bf r}) w[\Omega({\bf r})]
\label{2.9}
\end{equation}
is conserved, $\partial_t I_w = 0$. All such conservation laws may be conveniently summarized by the function
\begin{equation}
g(\sigma) = \int_D d{\bf r} h({\bf r}) \delta[\sigma - \Omega({\bf r})],
\label{2.10}
\end{equation}
which is then conserved for each value of $-\infty < \sigma < \infty$. One may recover any $I_w$ from $g(\sigma)$ in the form
\begin{equation}
I_w = \int d\sigma g(\sigma) w(\sigma).
\label{2.11}
\end{equation}

An important consequence of (\ref{2.9}) is that, choosing $w(\Omega) = \Omega$, one obtains
\begin{equation}
I_1 = \int_D d{\bf r} h\Omega = \int_D d{\bf r} (\omega+f)
= \int_{\partial D} {\bf v} \cdot d{\bf l} + \int_D d{\bf r} f.
\label{2.12}
\end{equation}
It follows that the total circulation is conserved. In a multiply connected domain, it can be shown that the individual circulations
\begin{equation}
\Gamma_l = \int_{\partial D_l} {\bf v} \cdot d{\bf l}
\label{2.13}
\end{equation}
about any connected component $\partial D_l$, $l = 1,2,\ldots,N_D$, of the boundary are conserved as well. These generate an additional $N_D-1$ independent conserved integrals that are not expressible in terms of $g(\sigma)$. We adopt the sign convention here that $\partial D_1$ is the outermost boundary, so the circulation integral direction on all other $\partial D_l$, $l \geq 2$, is opposite. In particular, the total circulation appearing in (\ref{2.12}) is given by
\begin{equation}
\Gamma = \int_{\partial D} {\bf v} \cdot d{\bf l} = \Gamma_1 - \sum_{l=2}^{N_D} \Gamma_l.
\label{2.14}
\end{equation}

\begin{figure}
\includegraphics[width=2.8in,viewport=190 190 530 370,clip]{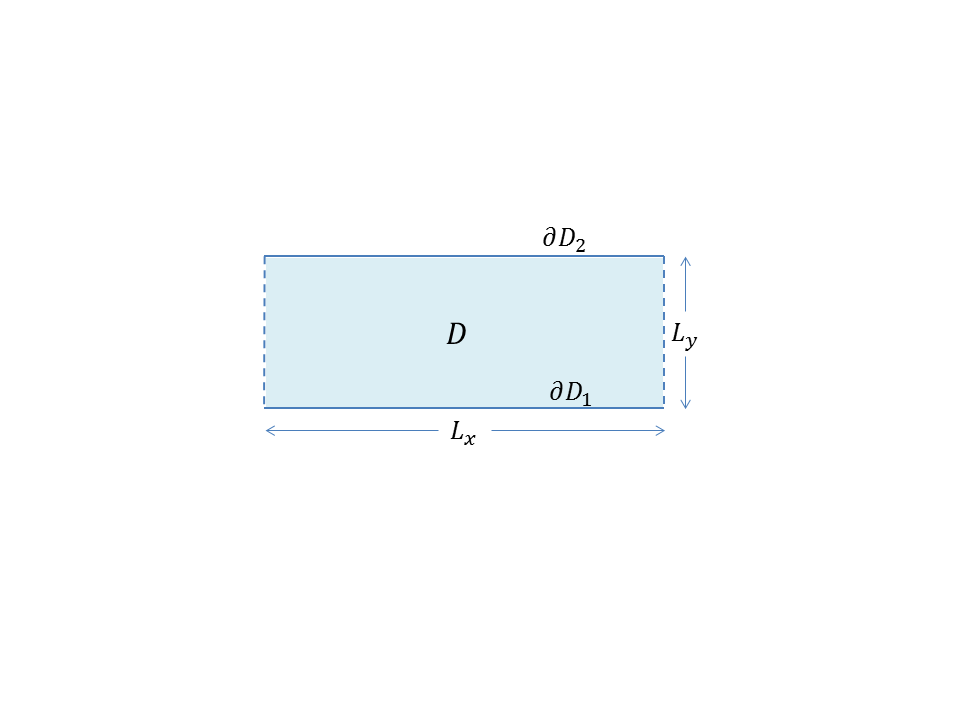}
\includegraphics[width=2.5in,viewport=190 110 500 440,clip]{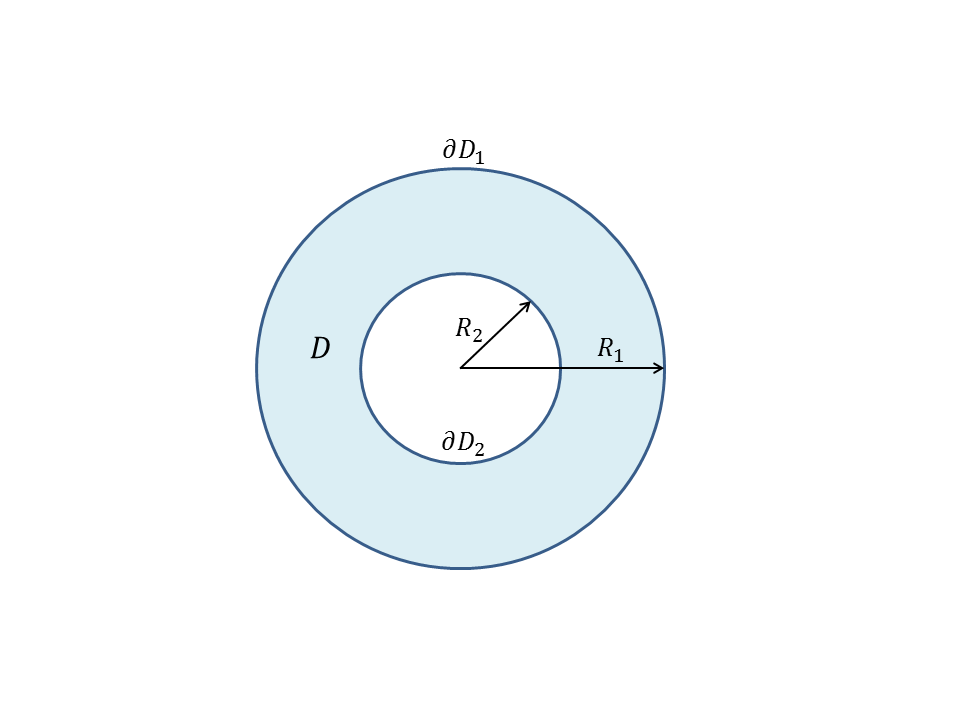}

\caption{\textbf{Top:} Translation invariant domain relevant to linear momentum conservation. Periodic boundary conditions are applied along $x$. \textbf{Bottom:} Rotation invariant domain relevant to angular momentum conservation.}

\label{fig:symdomains}
\end{figure}

\subsubsection{Energy and momentum}
\label{subsec:energy}

The conserved energy is a sum of kinetic and potential contributions:
\begin{equation}
E = \frac{\rho_0}{2} \int_D d{\bf r}
\left[h({\bf r}) |{\bf v}({\bf r})|^2
+ g \eta({\bf r})^2 \right].
\label{2.15}
\end{equation}

The canonical linear momentum is given by
\begin{equation}
{\bf P} = \rho_0 \int_D d{\bf r} \, h[{\bf v} + {\bf A}],
\label{2.16}
\end{equation}
where the vector potential is defined by $f = \nabla \times {\bf A}$ \cite{foot:vecpot}. If the system is translation invariant along a direction which we call ${\bf \hat x}$ (including the case of periodic boundary conditions along this direction, illustrated in the upper panel of Fig.\ \ref{fig:symdomains}), the momentum component $P_x = {\bf P} \cdot {\bf \hat x} $ is conserved. More explicitly, if $f = f(y)$ and $h_b = h_b(y)$ depend only on the orthogonal coordinate $y$, one may choose ${\bf A} = -F(y) {\bf \hat x}$ where $\partial_y F = f$, and one obtains the conserved integral
\begin{equation}
P_x = \rho_0 \int_D d{\bf r} h({\bf r}) [v_x({\bf r}) - F(y)].
\label{2.17}
\end{equation}
Using (\ref{2.1}), and judicious application of integration by parts and the boundary conditions, it is straightforward to verify directly that $\partial_t P_x = 0$.

The translation symmetry corresponds to the following Galilean transformation of the fields themselves:
\begin{eqnarray}
{\bf \bar v}({\bf r},t) &=& {\bf v}({\bf r} - {\bf \hat x} v_0 t,t)
+ v_0 {\bf \hat x}
\nonumber \\
\bar h({\bf r},t) &=& h({\bf r} - {\bf \hat x} v_0 t,t)
\nonumber \\
\bar \omega({\bf r},t) &=& \omega({\bf r} - {\bf \hat x} v_0 t,t)
\nonumber \\
\bar h_b(y) &=& h_b(y) - \frac{v_0}{g} F(y),
\label{2.18}
\end{eqnarray}
Thus, the same flow pattern boosted by an arbitrary velocity $v_0$ is a solution to (\ref{2.2}) if one imposes an additional bottom tilt proportional to $F(y)$. In the magnetic analogy \cite{foot:vecpot}, the latter corresponds to a ``Hall voltage'' that compensates for the change in Coriolis force induced by the change in the mean flow.

Similarly, in the presence of a rotational symmetry (circular or annular domain, illustrated in the lower panel of Fig.\ \ref{fig:symdomains})), the canonical angular momentum
\begin{equation}
L = \rho_0 \int_D d{\bf r} \, h {\bf r} \times ({\bf v} + {\bf A})
\label{2.19}
\end{equation}
is conserved. Here, the 2D vector cross product produces the scalar quantity ${\bf r} \times {\bf j} = x j_y - y j_x$. In this case $f = f(r)$ and  $h_b = h_b(r)$ depend only on the radial coordinate, and one may choose azimuthal ${\bf A} = \hat {\bm \theta} F(r)$, with $f = r^{-1} \partial_r(rF)$ to obtain the explicit form
\begin{equation}
L = \rho_0 \int_D d{\bf r} h({\bf r}) [{\bf r} \times {\bf v}({\bf r}) + rF(r)].
\label{2.20}
\end{equation}
It is again straightforward to verify directly that $\partial_t L = 0$.

The field symmetry corresponding to (\ref{2.20}) is the rotational Galilean transformation
\begin{eqnarray}
{\bf \bar v}({\bf r},t) &=& {\bf \hat R}_{\omega_0 t}
{\bf v}({\bf \hat R}_{-\omega_0 t}{\bf r},t)
+ \omega_0 r \hat {\bm \theta}
\nonumber \\
\bar h({\bf r},t) &=& h({\bf \hat R}_{-\omega_0 t}{\bf r},t)
\nonumber \\
\bar \omega({\bf r},t) &=& \omega({\bf \hat R}_{-\omega_0 t}{\bf r},t) + 2\omega_0
\nonumber \\
\bar h_b(r) &=& h_b(r) + \frac{\omega_0}{g} rF(r) - \frac{\omega_0^2}{2g} r^2
\nonumber \\
\bar f(r) &=& f(r) - 2\omega_0,
\label{2.21}
\end{eqnarray}
where $\hat {\bm \theta} = {\bf \hat z} \times {\bf \hat r}$ is the azimuthal unit vector, and ${\bf \hat R}_\alpha {\bf r} = r [\cos(\alpha){\bf \hat r} + \sin(\alpha) \hat {\bm \theta}]$ applies the 2D rotation by angle $\alpha$. In this case, the transformation preserves the identical flow pattern, but it now undergoes a net rotation at arbitrary angular rate $\omega_0$, and is maintained by both a bottom tilt correction [this time including also a centrifugal potential $\propto (\omega_0 r)^2$] and a change in the Coriolis parameter itself.

\section{Expressions in terms of canonical fields $\Omega,Q,h$}
\label{sec:canonfields}

Given the fundamental role of the conservation laws in the statistical mechanical treatment, it is useful to express quantities in terms of $\Omega$ and the compressional part of the velocity field
\begin{equation}
Q = \frac{q}{h},\ \ q \equiv \nabla \cdot {\bf v}.
\label{3.1}
\end{equation}
To this end, one decomposes ${\bf v}$ into rotational and compressional components:
\begin{equation}
{\bf v} = \nabla \times \psi - \nabla \phi,
\label{3.2}
\end{equation}
where the 2D curl of a scalar is defined by $\nabla \times \psi = (\partial_y \psi, -\partial_x \psi)$. Both terms are chosen transverse to any free-slip boundary: ${\bf \hat l} \cdot \nabla \psi = 0$, ${\bf \hat n} \cdot \nabla \phi = 0$, where ${\bf \hat l}$ and ${\bf \hat n}$ are the boundary tangent and normal unit vectors, respectively. Both obey any periodic boundary condition that might present as well. Substituting the form (\ref{3.2}) into (\ref{2.6}) and (\ref{3.1}), one obtains
\begin{equation}
\left[\begin{array}{c}
\omega \\ q
\end{array} \right]
= \left[\begin{array}{c}
h\Omega - f \\ hQ
\end{array} \right]
= -\nabla^2
\left[\begin{array}{c}
\psi \\ \phi
\end{array} \right].
\label{3.3}
\end{equation}

\begin{figure}
\includegraphics[width=2.5in,viewport=200 160 515 400,clip]{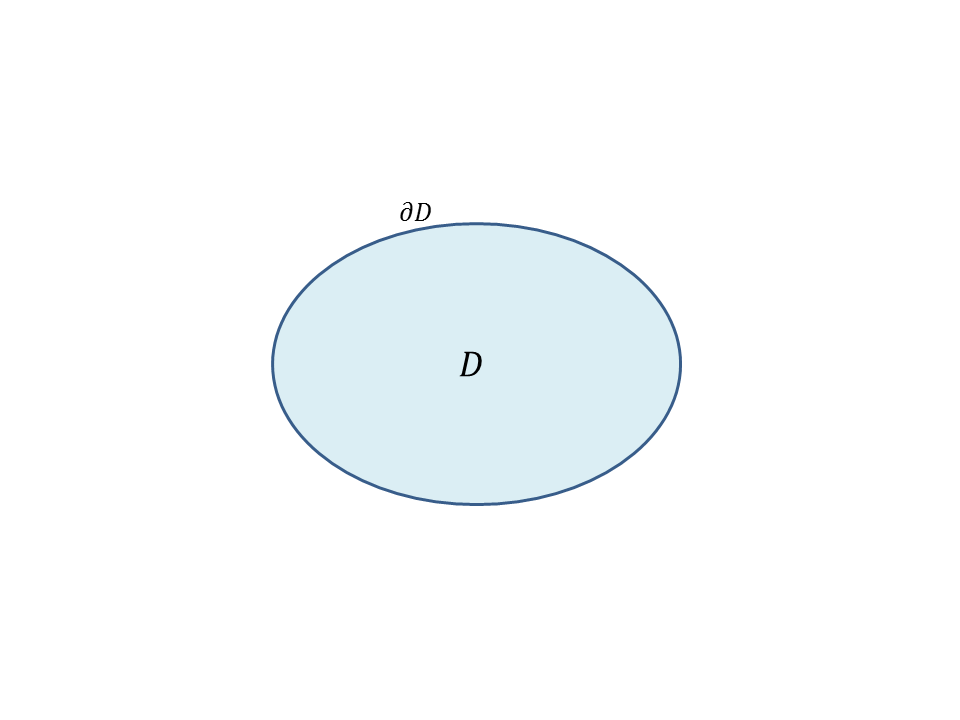}

\bigskip

\includegraphics[width=2.5in,viewport=200 180 520 410,clip]{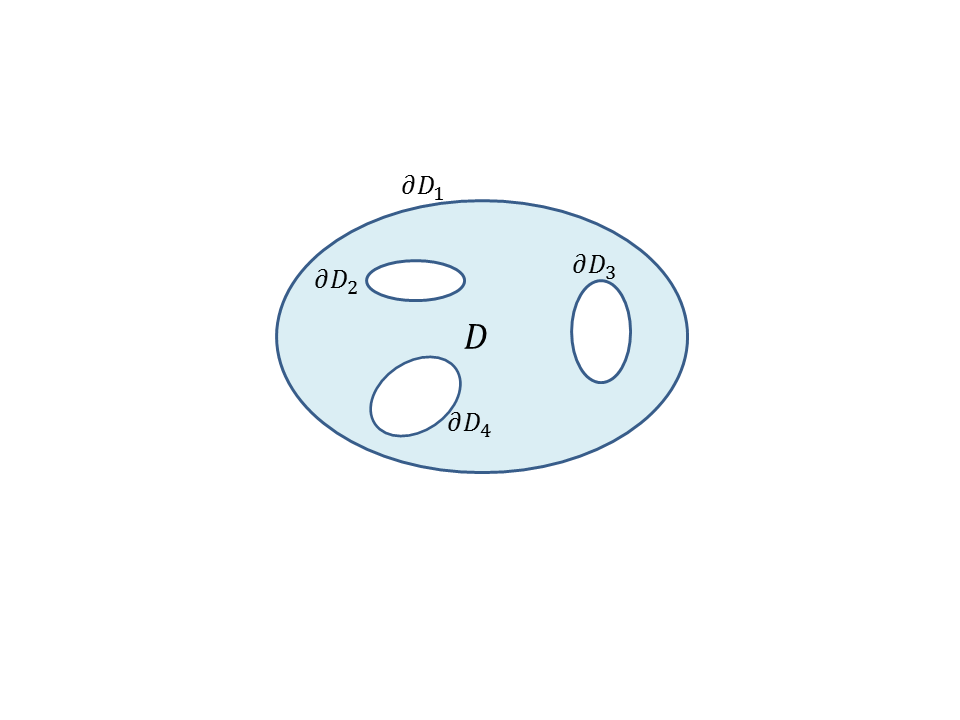}

\caption{\textbf{Top:} Simply connected domain. \textbf{Bottom:} Multiply connected domain. Independent circulation integrals $\Gamma_l$ and stream function values $\psi^0_l$ are associated with each internal boundary $\partial D_l$, and are related via (\ref{3.13}) through the potential flow circulations (\ref{3.11}).}

\label{fig:simmultdomains}
\end{figure}

\subsection{Potential and non-potential flow decomposition}
\label{sec:vortpot}

The free-slip condition on $\psi$ implies that it is constant on each connected component of the boundary. For a simply connected domain (top panel of Fig.\ \ref{fig:simmultdomains}), one may specify the Dirichlet condition $\psi|_{\partial D} \equiv 0$. However, for a multiply connected domain (bottom panel of Fig.\ \ref{fig:simmultdomains}), one has the interesting consequence that the boundary value differences of $\psi$ can fluctuate. Multiply connected domains are emphasized here because they play a key role in the presence of conserved momenta. To account for these conservation laws in the statistical mechanics treatment, one must separate out the corresponding \emph{potential flow} contributions.

To account for this dynamical degree of freedom we write $\psi$ in the form of a superposition:
\begin{equation}
\psi({\bf r}) = \psi^V({\bf r}) + \psi^P({\bf r}),
\label{3.4}
\end{equation}
in which the vortical component $\psi^V$ \emph{vanishes} on every free slip boundary component, and contains all contributions to $\omega$,
\begin{equation}
-\nabla^2 \psi^V = \omega
\label{3.5}
\end{equation}
while the ``potential flow'' field $\psi^P$ matches the boundary values of $\psi$,
\begin{equation}
\psi^P|_{\partial D} = \psi|_{\partial D},
\label{3.6}
\end{equation}
while producing zero circulation and compression:
\begin{equation}
-\nabla^2 \psi^P = 0 \ \ \Leftrightarrow \ \
\nabla \times {\bf v}^P
= 0 = \nabla \cdot {\bf v}^P,
\label{3.7}
\end{equation}
where we define
\begin{equation}
{\bf v}^P = \nabla \times \psi^P,\ \
{\bf v}^V = \nabla \times \psi^V,\ \
{\bf v}^C = -\nabla \phi.
\label{3.8}
\end{equation}
The orthogonality conditions
\begin{equation}
\int_D d{\bf r} {\bf v}^I \cdot {\bf v}^J = 0,
\label{3.9}
\end{equation}
for $I \neq J = P,V,C$ follows through integration by parts, and the use of the boundary conditions.

Since both $\psi^V$ and $\phi$ satisfy homogeneous boundary conditions, one obtains the inverse relations
\begin{eqnarray}
\psi^V({\bf r}) &=& \int_D d{\bf r}' G_D({\bf r},{\bf r}') \omega({\bf r}')
\nonumber \\
\phi({\bf r}) &=& \int_D d{\bf r}' G_N({\bf r},{\bf r}') q({\bf r}'),
\label{3.10}
\end{eqnarray}
in which $G_D$ and $G_N$ are, respectively, the Dirichlet and Neumann Green functions of the Laplacian for the domain $D$.

The aim in what follows is to show that both $\psi^V$ and $\psi^P$ are fully determined by $\omega$ and the conserved circulations (\ref{2.13}). To this end, we further decompose
\begin{equation}
\psi^P({\bf r}) = \psi^0_1 + \sum_{l=2}^{N_D} (\psi^0_l - \psi^0_1) \psi^P_l({\bf r}),
\label{3.11}
\end{equation}
in which $\psi^0_l = \psi|_{\partial D_l}$ is the value of $\psi$ on connected boundary component $\partial D_l$, $l=1,2,\ldots,N_D$, and the ``potential flow eigenfunctions'' are independent solutions to the Laplace equation on $D$ obeying
\begin{equation}
\psi^P_l({\bf r})|_{\partial D_m} = \delta_{lm},\
l=2,3,\ldots,N_D,
\label{3.12}
\end{equation}
i.e., the boundary value is nonzero only on the matching boundary component. We define as well the symmetric, positive definite array of inner products
\begin{equation}
\Gamma^P_{lm} = \int_D d{\bf r} {\bf v}_l^P \cdot {\bf v}_m^P
= \int_{\partial D_l} {\bf v}_m^P \cdot d{\bf l}
= \int_{\partial D_m} {\bf v}_l^P \cdot d{\bf l},
\label{3.13}
\end{equation}
in which the boundary integrals follow by substituting ${\bf v}^P_l = \nabla \times \psi^P_l$, integrating by parts, and using (\ref{3.9}).

The potential eigenfunctions may also be used to decompose the circulation integrals (\ref{2.13}) into potential and vortex contributions (the contribution from the compressional component $\phi$ trivially vanishes). Through integration by parts, and recalling the sign convention (\ref{2.14}), it is easy to check that
\begin{eqnarray}
\int_D d{\bf r} \psi^P_l({\bf r}) \omega({\bf r})
&=& \int_{\partial D} \psi^P_l {\bf v} \cdot d{\bf l}
+ \int_D d{\bf r} {\bf v} \cdot {\bf v}^P_l
\nonumber \\
&=& -\Gamma_l + \sum_{m=2}^{N_D} \Gamma^P_{lm}
(\psi^0_m - \psi^0_1).\ \ \ \ \ \
\label{3.14}
\end{eqnarray}
It follows that the conserved circulation integrals (\ref{2.13}) may be decomposed in the form
\begin{eqnarray}
\Gamma_l &=& \Gamma^V_l + \Gamma^P_l
\nonumber \\
\Gamma^V_l &=& \int_{\partial D_l} {\bf v}^V \cdot d{\bf l}
 = -\int_D d{\bf r} \psi^P_l({\bf r}) \omega({\bf r})
\nonumber \\
\Gamma^P_l &=& \int_{\partial D_l} {\bf v}^P \cdot d{\bf l}
= \sum_{m=2}^{N_D} \Gamma^P_{lm} (\psi^0_m - \psi^0_1).
\label{3.15}
\end{eqnarray}
This leads to the interpretation of $\psi^P({\bf r})$ as the circulation about boundary component $l$ due to a unit point vortex at ${\bf r}$. One obtains, in particular,
\begin{equation}
\psi^0_l - \psi^0_1 = \sum_{l=2}^{N_D}
[\Gamma^P]^{-1}_{lm} (\Gamma_m - \Gamma^V_m),
\label{3.16}
\end{equation}
demonstrating, as required, that the inhomogeneous boundary values, though fluctuating with the flow, are in fact fully specified by the vorticity field and the conserved integrals.

\subsubsection{Periodic strip geometry}
\label{subsec:periodicstrip}

Relevant to systems with linear momentum conservation (\ref{2.17}), the two connected boundary components are the lower and upper boundaries, $y_1 < y_2$, of the periodic strip of length $L_x$ (top panel of Fig.\ \ref{fig:symdomains}). There is a single potential flow eigenfunction, representing uniform flow along the channel:
\begin{equation}
\psi^P_2({\bf r}) = \frac{y-y_1}{L_y},\ \
{\bf v}^P_2 = \frac{1}{L_y} {\bf \hat x},
\label{3.17}
\end{equation}
where $L_y = y_2 = y_1$. The circulation integral follows in the form
\begin{equation}
\Gamma^P \equiv \Gamma^P_{22} = \frac{L_x}{L_y}.
\label{3.18}
\end{equation}

Well known analytic series forms for the Green functions $G_N,G_D$ entering (\ref{3.10}) may be derived using the method of images.

\subsubsection{Annular geometry}
\label{subsec:annulus}

Relevant to systems with angular momentum conservation, for an annular geometry, with inner and outer radii $0 \leq R_2 < R_1$ (lower panel of Fig.\ \ref{fig:symdomains}), the single eigenfunction corresponds to the axial flow
\begin{equation}
\psi^P_2({\bf r}) = \frac{\ln(r/R_1)}{\ln(R_2/R_1)},\ \
{\bf v}^P_2 = \frac{1}{\ln(R_2/R_1) r} \hat {\bm \theta}.
\label{3.19}
\end{equation}
The circulation integral takes the form
\begin{equation}
\Gamma^P = \frac{2\pi}{\ln(R_2/R_1)}.
\label{3.20}
\end{equation}

Once again, well known analytic series forms for the Green functions in (\ref{3.10}) may be derived in polar coordinates.

\subsection{Kinetic energy}
\label{sec:ke}

The substitution of the decomposition ${\bf v} = {\bf v}^V + {\bf v}^C + {\bf v}^P$, along with the representations (\ref{3.8}), (\ref{3.10}) and (\ref{3.11}), allows one to express into the kinetic part of the energy (\ref{2.15}),
\begin{equation}
E_K = \frac{\rho_0}{2} \int_D d{\bf r} h({\bf r}) |{\bf v}({\bf r})|^2,
\label{3.21}
\end{equation}
as a nonlocal quadratic functional of $\omega,q,\psi_l^0$, including also $h$. Note that in the periodic strip or the annulus, there is only a single term in the sum (\ref{3.11}), $l = m = N_D = 2$. Only $\Gamma^P_{22}$ enters (\ref{3.15}), given by the explicit forms (\ref{3.18}) or (\ref{3.20}), respectively. Substituting $\omega = h\Omega - f$ and $q = hQ$, as well as (\ref{3.16}), provides the explicit representation in terms of the basic fields $\Omega,Q,h$. The result is quite messy, including nonvanishing cross-terms, despite (\ref{3.9}), due to presence of $h$. This expression is not actually needed in the analysis below, but for completeness is written out in App.\ \ref{app:KEPi}.

\subsection{Conserved momenta}
\label{sec:consmomenta}

The kinetic parts of the linear and angular momenta, (\ref{2.17}) and (\ref{2.20}), may similarly be decomposed into vortical, compressional, and potential components. It is useful to write these in the form
\begin{eqnarray}
\Pi &=& \Pi_K + \Pi_h
\nonumber \\
\Pi_K &=& \rho_0 \int_D d{\bf r} h({\bf r})
{\bf v}_\Pi({\bf r}) \cdot {\bf v}({\bf r})
\nonumber \\
\Pi_h &=& \rho_0 \int_D d{\bf r} h({\bf r}) F_\Pi({\bf r})
\label{3.22}
\end{eqnarray}
where
\begin{eqnarray}
{\bf v}_\Pi({\bf r}) &=& \left\{\begin{array}{ll}
\hat {\bf x}, & \Pi = P_x \\
r \hat {\bm \theta}, & \Pi = L
\end{array}  \right.
\nonumber \\
F_\Pi({\bf r}) &=& \left\{\begin{array}{ll}
-F(y), & \Pi = P_x \\
r F(r), & \Pi = L.
\end{array} \right.
\label{3.23}
\end{eqnarray}
Substituting the decomposition (\ref{3.8}) of ${\bf v}$, $\Pi_K$ may be written out as a linear functional of $\omega,q,\psi^0_l$, depending also nonlocally on $h$. These expressions, given in App.\ \ref{app:KEPi}, will again not actually be needed below.

\subsection{Example: flat-bottom Euler equation}
\label{sec:eulereg}

The Euler equation on a flat bottom is obtained by setting $h = H_0$, $\eta = h_b = 0$, $\nabla \cdot {\bf v} = 0$, hence ${\bf v} = \nabla \times \psi$, and $\phi = 0$ (no compressional component). The potential flow eigenfunction expansion (\ref{3.11}) remains exactly as before.

The vortex contribution to the stream function is still given by first line of (\ref{3.10}), and the vortex contribution to the kinetic energy $E_K = E_K^V + E_K^P$ follows in the familiar Coulomb-like form
\begin{equation}
E_K^V = \frac{1}{2} \rho_0 H_0 \int_D d{\bf r} \int_D d{\bf r}'
\omega({\bf r}) G_D({\bf r},{\bf r}') \omega({\bf r}').
\label{3.24}
\end{equation}
Since $\Gamma_{lm}[h] = H_0 \Gamma_{lm}^P$, the potential flow contribution is given by
\begin{equation}
E_K^P = \frac{1}{2} \rho_0 H_0 \sum_{l,m = 2}^{N_D}
\Gamma_{lm}^P (\psi^0_l - \psi^0_1)(\psi^0_m - \psi^0_1)
\label{3.25}
\end{equation}
The cross term vanishes by orthogonality (\ref{3.9}).

With linear momentum conservation on a periodic strip, one obtains from (\ref{2.16}) the form
\begin{equation}
P_x = \rho_0 (v_0 - v_f) V_D,
\label{3.26}
\end{equation}
where $V_D = H_0 A_D$ is the system volume,
\begin{equation}
v_0 = \frac{\psi^0_2 - \psi^0_1}{L_y}
\label{3.27}
\end{equation}
is the (conserved) mean flow speed along the periodic dimension, $L_y = y_2-y_1$ is the strip width, and
\begin{equation}
v_f = \frac{1}{L_y} \int_{y_1}^{y_2} F(y) dy
\label{3.28}
\end{equation}
is a speed defined by the Coriolis effect. The momentum resides entirely in the potential component of the flow in this case, and the boundary values $\psi_{1,2}^0$ are both conserved. In particular, the value of $P_x$ fully specifies the boundary conditions and the energy in the potential flow. It fully specifies the potential contribution to the kinetic energy as well:
\begin{equation}
E_K^P = \frac{1}{2} \rho_0 H_0 \Gamma^P (\psi^0_2 - \psi^0_1)^2
= \frac{1}{2} \rho_0 V_D v_0^2.
\label{3.29}
\end{equation}

The circulation integral $\Gamma_2 = \Gamma_2^V + \Gamma^P (\psi^0_2 - \psi^0_1)$ follows directly from (\ref{3.15}). Inserting (\ref{3.17}), one sees that the vorticity contribution
\begin{equation}
\Gamma^V_2 = - \frac{1}{L_y} \int_D d{\bf r} (y-y_1) \omega({\bf r})
\label{3.30}
\end{equation}
is separately conserved, and also equivalent to momentum conservation.

For the annular geometry, one may express
\begin{eqnarray}
\int_D d{\bf r} {\bf r} \times {\bf v}
&=& -\frac{1}{2} \int_D d{\bf r}
{\bf v}({\bf r}) \cdot \nabla \times (r^2-R_1^2)
\label{3.31} \\
&=& \frac{1}{2} (R_1^2 - R_2^2) \Gamma_2
- \frac{1}{2} \int_D d{\bf r} (r^2 - R_1^2) \omega.
\nonumber
\end{eqnarray}
The angular momentum may therefore be written in the form
\begin{eqnarray}
L &=& \rho_0 H_0 \left[L_2 + \frac{1}{2}(R_1^2 - R_2^2) \Gamma_2
+ F_2 \right]
\nonumber \\
L_2 &=& \frac{1}{2} \int_D d{\bf r} (R_1^2 - r^2) \omega({\bf r})
\nonumber \\
F_2 &=& 2\pi \int_{R_2}^{R_1} r^2 F(r) dr,
\label{3.32}
\end{eqnarray}
which expresses it entirely in terms of the vorticity field and the conserved boundary circulations. Conservation of $L$ therefore produces the new conserved vorticity second moment $L_2$, analogous to the first moment (\ref{3.30}).

The potential flow is equivalent to a point vortex at the origin, and one obtains
\begin{equation}
E_K^P = \frac{\pi (\psi^0_2 - \psi^0_1)^2}{\rho_0 H_0 \ln(R_1/R_2)}.
\label{3.33}
\end{equation}
Using (\ref{3.15}), (\ref{3.19}) and (\ref{3.20}), the vorticity contribution to the circulation integral is
\begin{equation}
\Gamma_2^V = \int_D d{\bf r} \frac{\ln(r/R_1)}{\ln(R_1/R_2)} \omega({\bf r})
\label{3.34}
\end{equation}

Unlike for the linear momentum case, $\Gamma_2^V$, along with the boundary value $\psi_2^0 - \psi_1^0$, is not conserved, hence fluctuates with the flow. The reason for the difference is that in the linear momentum case ${\bf v}_{P_x}({\bf r}) = \hat {\bf x} = L_y {\bf v}^P_2({\bf r})$ happens to coincide with the potential eigenfunction, whereas ${\bf v}_L({\bf r}) = r \hat {\bm \theta}$ is distinct from ${\bf v}^P_2 \propto \hat {\bm \theta}/r$. In particular, the former has nonzero vorticity $\omega_L = 2$.

\section{Fluid system statistical mechanics}
\label{sec:statmech}

We seek a description of the equilibrium flows of the shallow water system, with conserved integrals defined by the energy (\ref{2.15}), advection constraints (Casimirs) (\ref{2.10}), the circulation integrals (\ref{2.13}), and momentum (\ref{2.17}) or (\ref{2.20}), if present. The equilibrium phase space measure $d\nu(\Gamma) = \rho(\Gamma) d\Gamma$, and the Liouville theorem from which it follows, are described in detail in App.\ \ref{app:liouville}. We work in the grand canonical ensemble with phase space probability density
\begin{equation}
\rho = \frac{1}{Z} e^{-\beta {\cal K}},\
Z \equiv \int d\Gamma e^{-\beta {\cal K}}
\label{4.1}
\end{equation}
and generalized Hamiltonian
\begin{equation}
{\cal K}[h,{\bf v}] = E - \alpha \Pi
- \sum_{l=2}^{N_D} \gamma_l \Gamma_l
- \int_D d{\bf r} h({\bf r}) \mu[\Omega({\bf r})].
\label{4.2}
\end{equation}
The function $\mu(\sigma)$ is the Lagrange multiplier function conjugate to $g(\sigma)$, and $\Pi$ denotes the conserved momentum ($P_x$ or $L$), if present---see (\ref{3.22}). The objective is to use this form to compute the free energy density
\begin{equation}
{\cal F}[\beta,\alpha,{\bm \gamma},\mu]
= -\frac{1}{\beta A_D} \ln(Z),
\label{4.3}
\end{equation}
which characterizes the equilibrium state.

The phase space integral (\ref{4.1}) is a formal infinite-dimensional functional integral over all possible fluid field configurations, weighted by the density $\rho(\Gamma)$. In order to perform computations, a finite-dimensional approximation is first constructed by discretizing the domain $D$ using a finite mesh (for simplicity, here taken as a uniform square mesh), replacing ${\bf r} \to {\bf r}_i$ by a discrete index $i$. To make physical sense, the continuum limit, taken at the end, must produce a finite, well defined form for ${\cal F}$, and this requirement will enforce nontrivial scaling of some parameters, especially the temperature $T = 1/\beta$.

Given the prominent role played by the potential vorticity, we use the statistical measure (\ref{A25}), defined in terms of unrestricted integrals over each grid value of $(\Omega,Q,h)$, as well as the Dirichlet boundary values $\psi_l^0$. The partition function takes the form
\begin{eqnarray}
Z &=& \prod_{l=2}^{N_D} \int d\psi^0_l \frac{\rho_0}{P_0}
\prod_i \int h_i^4 dh_i
\frac{\rho_0^2 \Delta x^2}{H_0 P_0^2}
\nonumber \\
&&\times\ \int dQ_i d\Omega_i
e^{-\beta {\cal K}[\Omega,Q,h,{\bm \psi}^0]}
\label{4.4}
\end{eqnarray}
where $\Delta x \to 0$ is the mesh size, and, as discussed in App.\ \ref{app:liouville}, the constant factors $\rho_0/P_0$ and $\rho_0^2/H_0 P_0^2$ are introduced for convenience to make the partition function dimensionless ($P_0$ has dimensions of momentum or mass current density ${\bf j}$).

\subsection{Form of generalized Hamiltonian}
\label{sec:formgenham}

In a symmetric domain, the $\alpha \Pi$ term is present, and some manipulations are required to put the combination $E - \alpha \Pi$ into a convenient form. By completing the square in various terms one obtains
\begin{eqnarray}
F &\equiv& E - \alpha \Pi = F^v + F^h + F^0
\nonumber \\
F^v &=& \frac{1}{2} \int d{\bf r} h \left|{\bf v} - \alpha {\bf v}_\Pi \right|^2
\nonumber \\
F^h &=& \frac{1}{2} \rho_0 g \int_D d{\bf r} \bar \eta^2
\nonumber \\
F^0 &=& -\rho_0 g \int_D d{\bf r}
\left[(h_b-H_0) \delta h_b + \frac{1}{2} \delta h_b^2 \right]
\label{4.5}
\end{eqnarray}
in which we define
\begin{eqnarray}
\bar \eta &=& h + \bar h_b - H_0
\nonumber \\
\bar h_b &=& h_b + \delta h_b
\nonumber \\
\delta h_b &=& -\frac{\alpha}{g}
\left(F_\Pi + \frac{1}{2} \alpha |{\bf v}_\Pi|^2 \right).
\label{4.6}
\end{eqnarray}
The term $F^0(\alpha)$ is constant, but does depend on the Lagrange multiplier $\alpha$. Note that only the full velocity ${\bf v}$ appears: the decompositions (\ref{3.2}) and (\ref{3.8}) will be exploited at a later step.

One may write the vorticity combination
\begin{eqnarray}
\omega - \alpha \omega_\Pi &=& hQ - \bar f
\nonumber \\
\bar f({\bf r}) &=& f({\bf r}) + \alpha \omega_\Pi({\bf r})
\nonumber \\
&=& \left\{\begin{array}{ll}
f(y), & \Pi = P_x \\
f(r) + 2 \alpha, & \Pi = L.
\end{array} \right.
\label{4.7}
\end{eqnarray}
The factor of $\omega_L = 2$ is obtained from (\ref{3.23}). The transformations (\ref{4.6}) and (\ref{4.7}) correspond precisely to the symmetry transformations (\ref{2.18}) and (\ref{2.21}) with $v_0 = -\alpha$ and $\omega_0 = -\alpha$, respectively. In this way, the $\alpha \Pi$ term effectively identifies the frame of reference in which the translation or rotation velocity vanishes.

\subsection{KHS transformation}
\label{sec:khs}

In order to simplify the calculation, we perform a Kac--Hubbard--Stratanovich (KHS) transformation by introducing an auxilliary Laplace transform 2D current density field ${\bf J}$. Its equilibrium average will eventually be related to the large scale flow. This field is used to convert the kinetic energy term into a term linear in the velocity via the Gaussian identity
\begin{equation}
e^{-\frac{1}{2} \beta \rho_0 \Delta x^2 h_i |{\bf V}_i|^2}
= \int_C d{\bf J}_i
\frac{e^{\beta \Delta x^2 |{\bf J}_i|^2/2 \rho_0 h_i}}
{2\pi \rho_0 h_i/\beta \Delta x^2}
e^{-\beta \Delta x^2 {\bf J}_i \cdot {\bf V}_i}
\label{4.8}
\end{equation}
applied independently to each site $i$, and used with ${\bf V} = {\bf v} - \alpha {\bf v}_\Pi$. The subscript $C$ is a complex integration contour, for each component of ${\bf J}_i$, that runs parallel to the imaginary axis. Here and below, for any 2D vector, we adopt the notation $|{\bf J}|^2 = {\bf J} \cdot {\bf J}$, which does \emph{not} include a complex magnitude.

The result (\ref{4.8}) holds for arbitrary real axis intersection point, but saddle point and other considerations will determine a convenient choice below. This identity is sensible in the limit $\Delta x \to 0$ only if the combination
\begin{equation}
\bar \beta = \beta \Delta x^2
\label{4.9}
\end{equation}
remains finite. Thus, the fluid hydrodynamic temperature $T = 1/\beta = \Delta x^2 \bar T$ (in contrast to the physical thermodynamic temperature) must \emph{vanish} in the continuum limit in order to obtain nontrivial macroscopic flows. The physical motivation for this scaling, which recognizes that large scale hydrodynamic flows cannot be in equilibrium with microscopic thermal fluctuations, has been discussed extensively in the literature, see, e.g., Ref.\ \cite{MWC1992}. The Gaussian integral also converges only if $\beta > 0$: as observed in \cite{RVB2016}, the inclusion of height fluctuations precludes the negative temperature states observed for the Euler equation \cite{M1990,RS1991,MWC1992}.

Inserting this identity for each $i$, one obtains
\begin{eqnarray}
Z &=& \prod_{l=2}^{N_D} d\psi^0_l \frac{\rho_0}{P_0}
\prod_i \int h_i^3 dh_i
\frac{\bar \beta \rho_0 \Delta x^2}{2\pi H_0 P_0^2}
\nonumber \\
&&\times\ \int_C d{\bf J}_i \int dQ_i d\Omega_i
e^{-\beta \tilde {\cal F}[{\bf J},h,{\bf v}]},
\label{4.10}
\end{eqnarray}
in which the free energy functional takes the continuum form
\begin{eqnarray}
\tilde {\cal F}[{\bf J},h,{\bf v}] &=& \int_D d{\bf r}
\left\{{\bf J}({\bf r}) \cdot
[{\bf v}({\bf r}) - \alpha {\bf v}_\Pi({\bf r})]
- \frac{|{\bf J}({\bf r})|^2}{2 \rho_0 h({\bf r})} \right\}
\nonumber \\
&&+\ F^h + F^0 - \sum_{l=2}^{N_D} \gamma_l \Gamma_l
- \int_D d{\bf r} h({\bf r}) \mu[\Omega({\bf r})]
\nonumber \\
\label{4.11}
\end{eqnarray}

We now reexpress the ${\bf J} \cdot {\bf V}$ term in terms of the canonical fields. Given that there is a component of ${\bf J}$ associated with each nonzero component of ${\bf v}$, it makes sense to enforce the free slip boundary condition ${\bf J} \cdot \hat {\bf n} = 0$ on ${\bf J}$ as well. It follows that one may apply the same decomposition (\ref{3.8}) to obtain
\begin{equation}
{\bf J} = \nabla \times \Psi - \nabla \Phi,\ \ \Psi = \Psi^V + \Psi^P.
\label{4.12}
\end{equation}
Substituting (\ref{3.8}), (\ref{3.9}), and (\ref{3.10}), one obtains
\begin{eqnarray}
&&\int_D d{\bf r} {\bf J} \cdot ({\bf v} - \alpha {\bf v}_\Pi)
= \int_D d{\bf r} [(h\Omega-\bar f) \Psi^V + h Q \Phi]
\nonumber \\
&&+\ \sum_{l=2}^{N_D} (\Psi_l^0 - \Psi_1^0)
\left[\sum_{m=2}^{N_D} \Gamma_{lm}^P (\psi_m^0 - \psi_1^0)
- \alpha \Gamma_{\Pi,l} \right] \ \ \ \ \ \
\label{4.13}
\end{eqnarray}
in which the momentum circulations (defined only for the case $N_D = 2$) are obtained from (\ref{3.23}) in the form
\begin{equation}
\Gamma_{\Pi,2} = \int_D d{\bf r} {\bf v}_\Pi \cdot {\bf v}^P_2
= \left\{\begin{array}{ll}
L_x, & \Pi = P_x \\
\frac{2\pi (R_1^2 - R_2^2)}{\ln(R_2/R_1)}, & \Pi = L
\end{array} \right.
\label{4.14}
\end{equation}
and one identifies the explicit forms
\begin{eqnarray}
\Psi^V({\bf r}) &=& \int d{\bf r}' G_D({\bf r},{\bf r}')
\nabla' \times {\bf J}({\bf r}')
\nonumber \\
\Phi({\bf r}) &=& \int d{\bf r}' G_N({\bf r},{\bf r}')
\nabla' \cdot {\bf J}({\bf r}').
\label{4.15}
\end{eqnarray}
The potential flow component is similarly decomposed in the form
\begin{equation}
\Psi^P({\bf r}) = \Psi^0_1
+ \sum_{l=2}^{N_D} (\Psi^0_l - \Psi^0_1) \psi^P_l({\bf r}).
\label{4.16}
\end{equation}

With these substitutions, and using the circulation representation (\ref{3.15}), the free energy functional takes the explicit form
\begin{widetext}
\begin{eqnarray}
\tilde {\cal F}[{\bf J},h,{\bf v}] &=& \int_D d{\bf r}
\left\{\frac{1}{2} \rho_0 g \bar \eta({\bf r})^2 - \bar f({\bf r})
\left[\Psi^V({\bf r}) + \Psi^{\bm \gamma}({\bf r}) \right]
- \frac{|{\bf J}({\bf r})|^2}{2 \rho_0 h({\bf r})}
\right\}
\nonumber \\
&&+\ \int_D d{\bf r} h({\bf r}) \left\{
\left[\Psi^V({\bf r}) + \Psi^{\bm \gamma}({\bf r}) \right] \Omega({\bf r})
+ \Phi({\bf r}) Q({\bf r}) - \mu[\Omega({\bf r})]\right\}
\nonumber \\
&&+\ \sum_{l,m=2}^{N_D} \Gamma_{lm}^P
(\Psi_l^0 - \Psi_1^0 - \gamma_l) (\psi_m^0 - \psi_1^0)
- \alpha \sum_{l=2}^{N_D} \Gamma_{\Pi,l} (\Psi_l^0 - \Psi_1^0 - \gamma_l)
+ \bar F^0(\alpha,{\bm \gamma}),
\label{4.17}
\end{eqnarray}
\end{widetext}
where we define
\begin{eqnarray}
\bar F^0(\alpha,{\bm \gamma}) &=& F^0(\alpha)
- \alpha \sum_{l=2}^{N_D} \gamma_l \Gamma^\Pi_l
\nonumber \\
\Gamma^\Pi_2 &\equiv& \int_{\partial D_l} {\bf v}_\Pi \cdot d{\bf l}
= \Gamma_{\Pi,2} - \int_D d{\bf r} \omega_\Pi \psi^P_2
\nonumber \\
&=& \left\{\begin{array}{ll}
L_x, & \Pi = P_x \\
2\pi R_2^2, & \Pi = L
\end{array} \right.
\nonumber \\
\Psi^{\bm \gamma}({\bf r}) &=& \sum_{l=2}^{N_D} \gamma_l \psi^P_l({\bf r}).
\label{4.18}
\end{eqnarray}

The form (\ref{4.17}) achieves the goal of being entirely local in $h,\Omega,Q$: for given ${\bf J}$, the statistical factor $e^{-\beta \tilde {\cal F}}$ can be expressed as an independent product over sites $i$, allowing the integration over $h_i,Q_i,\Omega_i,\psi^0_m$ to be carried out explicitly. To proceed, we note first that the only dependence on $Q$ is in the second line of (\ref{4.17}). Choosing $\Phi_i$ to be pure imaginary, the former may be integrated out to produce a factor
\begin{equation}
\prod_i 2\pi \delta(i \bar \beta h_i \Phi_i)
= \prod_i \frac{2\pi}{\bar \beta h_i} \delta(i\Phi_i).
\label{4.19}
\end{equation}
Directly analogous to the change of variable from ${\bf v}$ in (\ref{A23}) to $(\Omega,Q,{\bm \psi}^0)$ in (\ref{A24}), one may change variables ${\bf J} \to (\Psi,\Phi,{\bm \Psi}^0)$, with constant Jacobian $\Delta x^{-2N_E}$:
\begin{equation}
\prod_i \int_C d{\bf J}_i
= \prod_{l=2}^{N_D} \int_C d\Psi^0_l
\prod_i \int_C \frac{d\Phi_i d\Psi^V_i}{\Delta x^2}.
\label{4.20}
\end{equation}
In each case, $C$ is again a contour parallel to the imaginary axis. The result of the $\Phi$ integral is therefore to simply set
\begin{equation}
\Phi_i \equiv 0 \ \ \forall i
\label{4.21}
\end{equation}
in $\tilde {\cal F}$ \cite{foot:Qint}. The factor $\prod_i (\bar \beta h_i)^{-1}$ produced by (\ref{4.19}) encompasses the contribution to the free energy from the fluctuations in $Q$ that have now been fully integrated out.

Similarly, ${\bm \psi}^0$ appears only in the last term in (\ref{4.17}). Choosing the ${\bm \Psi}^0$ contours so that $\Psi^0_l - \Psi^0_1 - \gamma_l$ are all pure imaginary, the ${\bm \psi}^0$ integrals produce the factor
\begin{equation}
\frac{1}{\det(\Gamma^P)} \prod_{l=2}^{N_D}
\delta[i(\Psi^0_l - \Psi^0_1 - \gamma_l)].
\label{4.22}
\end{equation}
The result of the ${\bm \Psi}^0$ integrals is therefore to simply replace
\begin{equation}
\Psi^0_l - \Psi^0_1 = \gamma_l,\ l=2,3,\ldots,N_D.
\label{4.23}
\end{equation}
Using (\ref{4.12}), (\ref{4.16}) and (\ref{4.21}), one identifies
\begin{eqnarray}
\Psi({\bf r}) &=& \Psi^V({\bf r}) + \Psi^{\bm \gamma}({\bf r})
\nonumber \\
|{\bf J}({\bf r})|^2 &=& |\nabla \times \Psi({\bf r})|^2
= |\nabla \Psi({\bf r})|^2
\label{4.24}
\end{eqnarray}
The first line implies that the circulation Lagrange multipliers simply enforce the boundary conditions $\Psi|_{\partial D_l} = \gamma_l$. We reiterate, here and below, that $|\nabla \Psi|^2 = \nabla \Psi \cdot \nabla \Psi$ does \emph{not} include a complex magnitude.

The end result of eliminating $Q,\Phi,{\bm \Psi}^0,{\bm \psi}^0$ is the partially reduced free energy functional
\begin{eqnarray}
\hat {\cal F}[\Psi,h,\Omega] &=& \bar F^0 + \int_D d{\bf r}
\bigg\{\frac{1}{2} \rho_0 g \bar \eta({\bf r})^2
- \frac{|\nabla \Psi({\bf r})|^2}{2 \rho_0 h({\bf r})}
\nonumber \\
&&+\ [\omega({\bf r}) - \alpha \omega_\Pi({\bf r})] \Psi({\bf r})
- \mu[\Omega({\bf r})] \bigg \}.
\nonumber \\
\label{4.25}
\end{eqnarray}

\subsection{Final effective models}
\label{sec:modelfinal}

There are two ways to proceed in order to further reduce (\ref{4.25}), each providing a rather different (but obviously equivalent) view of the underlying physics. The first is to integrate out $h,\Omega$ to obtain an effective theory in terms of the stream function $\Psi$ alone. This yields an effective nonlinear elastic membrane interpretation. The second is to integrate out $\Psi$ to obtain a dual effective theory in terms of $\Omega,h$. This yields the generalized Coulomb system interpretation. The latter, which is now completely independent of the KHS field ${\bf J}$, could also have been obtained by integrating out $Q,{\bm \psi}$ from ${\cal K}$ in (\ref{4.4}). However, the intermediate KHS route actually provides the more transparent derivation. We derive both models in sequence.

\subsubsection{Nonlinear elastic membrane model}
\label{subsec:nonlinmembrane}

In order to handle the $h_i,\Omega_i$ integrals, we define a function $W$ of three scalar arguments by
\begin{eqnarray}
e^{\bar \beta W(\tau,h_0,\xi)}
&=& \frac{\rho_0}{P_0}
\int_0^\infty \lambda^2 d\lambda \int d\sigma
e^{\bar \beta \lambda[\mu(\sigma) - \sigma \tau]}
\nonumber \\
&&\ \ \ \ \ \ \times\
e^{-\frac{1}{2} \bar \beta
[\xi/\rho_0 \lambda + \rho_0 g(\lambda + h_0)^2]}.\ \ \ \ \
\label{4.26}
\end{eqnarray}
The factor $\lambda^2$ originates from the factor $\bar \beta h_i^3$ in (\ref{4.10}), divided the factor $\bar \beta h_i$ in (\ref{4.19}). The remaining factor $\rho_0/P_0$ makes the result dimensionless. We observe here again that (1) this function makes sense only if $\bar \beta$ (not $\beta$) is finite, and (2) that the $\lambda$-integral converges only if $\bar \beta, \xi > 0$.

Combining (\ref{4.20}), (\ref{4.23}), and (\ref{4.24}), the partition function may be put in the form
\begin{equation}
Z = \prod_i \int_C \frac{d\Psi^V_i}{P_0 H_0}
e^{-\beta {\cal F}[{\bm \Psi}]}
\equiv \int D[\Psi^V] e^{-\beta {\cal F}[{\bm \Psi}]}
\label{4.27}
\end{equation}
with fully reduced (continuum limit) free energy functional
\begin{widetext}
\begin{eqnarray}
{\cal F}[\Psi] = \bar F^0(\alpha,{\bm \gamma})
- \int_D d{\bf r} \left\{\bar f({\bf r})\Psi({\bf r})
+ W\left[\Psi({\bf r}), \bar h_b({\bf r}) - H_0,\
-|\nabla \Psi({\bf r})|^2 \right] \right\}.
\label{4.28}
\end{eqnarray}
\end{widetext}
The physical interpretation of this model is that of an inhomogeneous, nonlinear elastic fluctuating membrane (see Ref.\ \cite{W2012} for a similar analogy in the context of the theory of magnetohydrodynamic equilibria). In the limit $\beta = \bar \beta/\Delta x^2 \to \infty$, the dependence on $|\nabla \Psi({\bf r})|^2$ ensures that $\Psi$ is continuous, with  $\delta \Psi = \Psi - \Psi^\mathrm{eq} = O(\Delta x/\sqrt{\bar \beta})$ differing only microscopically from its (smooth) equilibrium average $\Psi^\mathrm{eq}({\bf r}) = \langle \Psi({\bf r}) \rangle$ \cite{foot:C}. However, it follows that $|\nabla \Psi({\bf r})|^2 = O(1/\bar \beta)$ is a finite random variable, varying on the microscale $\Delta x$. For non-gradient terms inside ${\cal F}$, one is therefore free to replace $\Psi \to \Psi^\mathrm{eq}$ (whose form must eventually be determined self-consistently), and the first two arguments of $W$ may then be viewed as smooth, deterministic functions of ${\bf r}$. However, the third argument remains a fluctuating field, contributing nontrivially to the functional integral.

If $W(\xi)$ were a slowly varying function of its third argument, on the scale $\bar T = 1/\bar \beta$, then $W(\xi) \simeq W(\xi_0) + \partial_\xi W(\xi_0) (\xi - \xi_0)$, where $\xi_0 = -|\nabla \Psi^\mathrm{eq}|^2$, and the membrane becomes linear (though still inhomogeneous), with effective local surface tension defined by $\partial_\xi W(\xi_0)$. However, with increasing $\bar T$ the linear approximation fails, the surface tension depends on $\xi$ itself, and the model becomes intrinsically nonlinear. One may understand this effect from the point of view of the original shallow water system. With increasing $\bar T$ the microscopic height fluctuations, correlated with the current density fluctuations $\nabla \times \Psi$, increase to the point where the height field excursions become comparable to $H_0$, and one exits the regime of linear surface waves. In this sense, the behavior here is significantly more complex than that found in the magnetohydrodynamic problem, where terms equivalent to $|\nabla \Psi|^2$ always enter the free energy functional linearly \cite{W2012}.

Since $|\nabla \Psi|^2$ varies by $O(1)$ on the lattice scale $\Delta x$, one might hope that it possesses only short range correlations. If this were true, one could independently integrate it out at each point ${\bf r}$ according to its single site-statistics, as we did the fields $\Omega,Q,h,\Phi$ in obtaining ${\cal F}[\Psi]$ from $\tilde {\cal F}[{\bf J},\Omega,Q,h]$. Unfortunately precisely the opposite is the case: the curl-free condition on $\nabla \Psi$, makes it highly correlated from site to site. For example, for the simplest, linear, homogenous model one obtains logarithmic correlations $\langle [\Psi({\bf r}) - \Psi({\bf r}')]^2 \rangle \sim \ln(|{\bf r}-{\bf r}'|/\Delta x)$. Correspondingly, one obtains macroscopic-scale dipole-like correlations of the current $\nabla \times \Psi$ \cite{W2012}. Thus, ${\cal F}$ generates a highly nontrivial, strongly correlated statistical model, with no simple analytic form for the free energy. In Sec.\ \ref{sec:furtherprops} we will consider limits in which the fluctuations are small, and in which more explicit analytic progress can be made.

If, for convenience one separates \cite{foot:C}
\begin{widetext}
\begin{eqnarray}
{\cal F}[\Psi] &=& {\cal F}[\Psi^\mathrm{eq}]
+ {\cal F}^\mathrm{fluct}[\Psi^\mathrm{eq},\delta \Psi]
\nonumber \\
{\cal F}[\Psi^\mathrm{eq},\delta \Psi]
&=& -\int_D d{\bf r} \left\{W\left[\Psi^\mathrm{eq},\bar h_b - H_0,
-|\nabla (\Psi^\mathrm{eq} + \delta \Psi)|^2 \right]
 - W\left[\Psi^\mathrm{eq},\bar h_b - H_0,
-|\nabla \Psi^\mathrm{eq}|^2 \right] \right\}
\label{4.29}
\end{eqnarray}
into static and fluctuating parts, then the equilibrium free energy takes the form
\begin{eqnarray}
F[\Psi^\mathrm{eq}] &=& {\cal F}[\Psi^\mathrm{eq}]
+ F^\mathrm{fluct}[\Psi^\mathrm{eq}]
\label{4.30}\\
F^\mathrm{fluct}[\Psi^\mathrm{eq}] &=& -\frac{1}{\beta}
\ln\left\{\int D[\delta \Psi] e^{-\beta {\cal F}^\mathrm{fluct}
[\Psi^\mathrm{eq},\delta \Psi]} \right\},
\nonumber
\end{eqnarray}
which explicitly exposes the ``mean field'' and fluctuating components. The self-consistent equation for the large-scale equilibrium flow follows by minimizing $F$:
\begin{equation}
\frac{\delta F}{\delta \Psi^\mathrm{eq}({\bf r})} = 0.
\label{4.31}
\end{equation}

The functional derivative (\ref{4.31}) may be conveniently evaluated by first defining an intermediate average over the fields $\Omega,h$ using the functional $W$:
\begin{eqnarray}
n_{\Omega,h}({\bf r},\sigma,\lambda)
&\equiv& \langle \delta[\Omega({\bf r}) - \sigma]
\delta[h({\bf r}) - \lambda] \rangle_W
\nonumber \\
&=& \frac{\frac{\rho_0}{P_0} \lambda^2 e^{\bar \beta \lambda
\left[\mu(\sigma) - \sigma \Psi({\bf r}) \right]}
e^{\frac{1}{2} \bar \beta
\left\{|\nabla \Psi({\bf r})|^2/\rho_0\lambda
- \rho_0 g[\lambda + \bar h_b({\bf r}) - H_0]^2 \right\}}}
{e^{\bar \beta W\left[\Psi({\bf r}), \, \bar h_b({\bf r})-H_0, \,
-|\nabla \Psi({\bf r})|^2 \right]}},
\label{4.32}
\end{eqnarray}
which may be interpreted as the probability density for potential vorticity and fluid height at the point ${\bf r}$, for a given fixed realization of the field $\Psi$. The $e^{\bar \beta W}$ denominator ensures that the distribution is normalized. This interpretation is most easily derived by following the identical sequence of integration steps to obtain the results (\ref{4.21}) and (\ref{4.23}), but in the integration over the $h$ and $\Omega$ fields (in advance of the $\Psi$ integration), the delta functions then produce (\ref{4.32}) in place of free integration result (\ref{4.26}). With this definition one obtains
\begin{eqnarray}
-\partial_\tau W
&=& \int_0^\infty \lambda d\lambda
\int d\sigma \sigma n_{\Omega,h}({\bf r},\sigma,\lambda)
= \langle h({\bf r}) \Omega({\bf r}) \rangle_W
= \langle \omega({\bf r}) \rangle_W + f({\bf r})
\nonumber \\
2\rho_0 \partial_\xi W
&=& \int_0^\infty \frac{d\lambda}{\lambda}
\int d\sigma n_{\Omega,h}({\bf r},\sigma,\lambda)
= \left\langle \frac{1}{h({\bf r})} \right\rangle_W
\nonumber \\
-\partial_{h_0} W &=& \rho_0 g \int_0^\infty d\lambda
(\lambda + \bar h_b - H_0)
\int d\sigma n_{\Omega,h}({\bf r},\sigma,\lambda)
= \rho_0 g \langle \bar \eta({\bf r}) \rangle_W,
\label{4.33}
\end{eqnarray}
\end{widetext}
and one may express (\ref{4.31}) in the form
\begin{eqnarray}
\nabla \times {\bf V}^\mathrm{eq}
&=& \langle h({\bf r}) \Omega({\bf r}) \rangle + f({\bf r})
= \langle \omega({\bf r}) \rangle
\nonumber \\
{\bf V}^\mathrm{eq} &\equiv& \left\langle
\frac{\nabla \times \Psi({\bf r})}{\rho_0 h({\bf r})} \right\rangle
+ \alpha {\bf v}_\Pi({\bf r}),
\label{4.34}
\end{eqnarray}
in which the averages now include $\Psi$:  $\langle \cdot \rangle \equiv \langle \langle \cdot \rangle_W \rangle_{\cal F}$. In the presence of momentum conservation, ${\bf V}^\mathrm{eq}$ is the instantaneous mean flow velocity seen in the laboratory frame, while ${\bf J} = \rho_0 \langle h ({\bf v} - \alpha {\bf v}_\Pi) \rangle$ is current density in the translating or rotating frame of reference (hence, generated by the net vorticity $\Delta \omega^\mathrm{eq} = \langle \omega \rangle - \alpha \omega_\Pi$). In the latter frame, the equilibrium flow is time-independent, obtained from the transformation (\ref{2.18}) or (\ref{2.21}), with $v_0 = -\alpha$ or $\omega_0 = -\alpha$, respectively. The incompressibility condition on ${\bf J}^\mathrm{eq}$ still allows, in general, a nonzero compressible velocity field component $\langle q \rangle = \nabla \cdot {\bf V}^\mathrm{eq}$.

The equilibrium form $\Psi$ depends on the Lagrange multipliers $\beta,\mu,\alpha,{\bm \gamma}$, which must then be tuned to obtain prescribed values of the conserved integrals. The latter may be derived as equilibrium averages in the form
\begin{widetext}
\begin{eqnarray}
g(\sigma) &=& - \frac{\delta {\cal F}}{\delta \mu(\sigma)}
= \langle h({\bf r}) \delta[\Omega({\bf r}) - \sigma] \rangle
= \int_D d{\bf r} \int_0^\infty \lambda d\lambda \,
\langle n_{\Omega,h}({\bf r},\sigma,\lambda) \rangle_{\cal F}
\nonumber \\
\Gamma_l &=& -\frac{\partial {\cal F}}{\partial \gamma_l}
= -\frac{\partial \bar F^0}{\partial \gamma_l}
+ 2 \int_{\partial D_l} \langle \partial_\xi W
(\nabla \times \Psi) \rangle \cdot d{\bf l}
= \int_{\partial D_l} \left\langle
\frac{{\bf J} + \rho_0 h \alpha {\bf v}_\Pi}{\rho_0 h}
\right \rangle \cdot d{\bf l}
\nonumber \\
&=& \int_{\partial D_l} {\bf V} \cdot d{\bf l}
= \sum_{m=2}^{N_D} \Gamma^P_{lm}
\langle \psi^0_l - \psi^0_1 \rangle
- \int_D d{\bf r} \psi^P_l({\bf r})
\langle \omega({\bf r}) \rangle
\nonumber \\
\Pi &=& -\frac{\partial {\cal F}}{\partial \alpha}
= -\frac{\partial \bar F^0}{\partial \alpha}
+ \int_D d{\bf r} \left\{\omega_\Pi \langle \Psi \rangle
- \frac{1}{g} \langle \partial_{h_0}W \rangle
[F_\Pi + \alpha |{\bf v}_\Pi|^2] \right\}
\nonumber \\
&=& \int_D d{\bf r} \left({\bf v}_\Pi \cdot
\langle {\bf J} + \rho_0 h \alpha {\bf v}_\Pi \rangle
+ \rho_0 \langle h \rangle F_\Pi \right)
= \rho_0 \int_D d{\bf r} \langle h({\bf r}) \rangle
\left[{\bf v}_\Pi({\bf r}) \cdot \tilde {\bf V}({\bf r})
+ F_\Pi({\bf r}) \right]
\nonumber \\
E &=& \left[\frac{\partial (\bar \beta {\cal F})}
{\partial \bar \beta}\right]_{\bar \beta \alpha,
\bar \beta \mu, \bar \beta {\bm \gamma}}
= \frac{1}{2} \int_D d{\bf r} \left\{
\left\langle \frac{|{\bf J}
+ \rho_0 h \alpha {\bf v}_\Pi|^2}{\rho_0 h} \right\rangle
+ \rho_0 g \langle \eta^2 \rangle \right\},
\label{4.35}
\end{eqnarray}
\end{widetext}
which correspond to averages of (\ref{2.10}), (\ref{3.15}), (\ref{3.22}), and (\ref{2.15}). In the last expression for $\Pi$, we define a somewhat different measure of the mean velocity field $\tilde {\bf V}$ by
\begin{equation}
\tilde {\bf V}({\bf r})
= \frac{\langle {\bf J} + \rho_0 h \alpha {\bf v}_\Pi \rangle}
{\rho_0 \langle h \rangle}.
\label{4.36}
\end{equation}
Clearly, $\tilde {\bf V}$ and ${\bf V}$ become equivalent if the fluctuations in $h$ are small. In the computation of $\Gamma_l$, only the surface term survives in the first line by virtue of (\ref{4.34}). The last expression for $\Gamma_l$ uses (\ref{3.15}) to alternatively express the average potential flow circulation $\langle \Gamma^P_l \rangle = \int_D d{\bf r} {\bf v}^P_l \cdot {\bf V}$ in terms of an average of the boundary values.

\subsubsection{Generalized Coulomb model}
\label{subsec:coulomb}

Alternatively, one may integrate out $\Psi$ from (\ref{4.25}). For fixed height field $h$ the integral is Gaussian, and one obtains
\begin{widetext}
\begin{eqnarray}
Z &=& \prod_i \int_0^\infty h_i^2 dh_i \frac{\rho_0}{P_0 H_0^{1/2}}
\int d\Omega_i \sqrt{\det(G_h)} e^{-\beta \hat K[\Omega,h]}
\nonumber \\
\hat {\cal K}[\Omega,h] &=& \frac{1}{2} \int_D d{\bf r} \int_D d{\bf r}'
[\omega({\bf r}) - \alpha \omega_\Pi({\bf r})] G_h({\bf r},{\bf r}')
[\omega({\bf r}) - \alpha \omega_\Pi({\bf r})]
+ \int_D d{\bf r} \left\{\frac{1}{2} \rho_0 g_0 \bar \eta({\bf r})^2
- h({\bf r}) \mu[\Omega({\bf r})] \right\}
\nonumber \\
&&-\ \sum_{l=2}^{N_D} \gamma_l \int_{\partial D_l}
\tilde {\bf v}  \cdot d{\bf l}
+ \frac{A_D}{2 \bar \beta}
\ln\left(\frac{2\pi P_0^2}{\rho_0 H_0} \bar \beta \right) + F^0(\alpha),
\label{4.37}
\end{eqnarray}
\end{widetext}
in which, as before, $\omega = h\Omega - f$, and the Green function $G_h$ is defined by
\begin{equation}
-\nabla \cdot \frac{1}{\rho_0 h({\bf r})}
\nabla G_h({\bf r},{\bf r}') = \delta({\bf r}-{\bf r}'),
\label{4.38}
\end{equation}
with Dirichlet boundary conditions on $\partial_D$. The form (\ref{4.37}) differs significantly from the form of the tensor Green function (\ref{C3}) before $Q$ is integrated out (compare, especially, the $\omega$-$\omega$ block of ${\cal \hat G}_h$). The $\det(G_h) \approx \prod_i h_i$ term comes from the normalization of the Gaussian integral and adjusts the phase space measure defining $\int D[h]$. The $\ln(\bar \beta)$ term generates the equipartition contribution $1/2\bar \beta = \bar T/2$ to the energy density coming the fluctuating $Q$-field that has been integrated out. The quantity
\begin{equation}
\tilde \Psi({\bf r}) = \int_D d{\bf r}' G_h({\bf r},{\bf r}')
[\omega({\bf r}') - \alpha \omega_\Pi({\bf r}')]
\label{4.39}
\end{equation}
obeys $\nabla \times (h^{-1} \nabla \times \tilde \Psi) = \omega - \alpha \omega_\Pi$, and
\begin{equation}
\tilde {\bf j}({\bf r}) \equiv
h({\bf r})[\tilde {\bf v}({\bf r})
- \alpha {\bf v}_\Pi({\bf r})]
= \nabla \times \tilde \Psi({\bf r})
\label{4.40}
\end{equation}
therefore represents the divergence free component of the current density. This also defines the quantity $\tilde {\bf v}$ appearing in the circulation term in (\ref{4.37}).

For constant $h \equiv H_0$, $G_h$ becomes the Dirichlet Coulomb potential, and the corresponding term in the free energy coincides with that for the Euler equation. The smoothness of this potential, together with the constrained fluctuations in $\omega$, produce an energy that is completely dominated by the large scale flow \cite{M1990,RS1991,MWC1992}. The energy may therefore be obtained by substituting $\langle \omega \rangle$ for $\omega$, and this in turn produces an exact variational form for the free energy. On the other hand, the presence of $1/h$ here, with $O(1)$ fluctuations on the scale $\Delta x$, produces a finite microscale fluctuation energy contribution: $G_h$ (as well as $\tilde \Psi$) is continuous, but its gradient fluctuates on scale $\Delta x$, and is highly correlated with $h$. It follows that one \emph{cannot} simply substitute $\langle \omega \rangle$ for $\omega$ and $\langle G_h \rangle$ for $G_h$. The reasons for this failure are equivalent to the correlated site-to-site fluctuations of $|\nabla \Psi|^2$ found in the membrane formulation (\ref{4.27}). One concludes again that the model free energy does not reduce to a variational mean field form.

\section{Simplifying limits and further properties of the models}
\label{sec:furtherprops}

In this section we consider simplifying limits in which more explicit computations can be carried out, and used these to explore further properties of the models. The critical assumption will be that the fluctuations are small, so that $\Psi \simeq \Psi^\mathrm{eq}$ may be treated as a fixed, nonfluctuating field.

\subsection{Variational limit}
\label{sec:varlimit}

The variational or mean field limit is defined by neglecting $F^\mathrm{fluct}$. In particular one sets $\Psi = \Psi^\mathrm{eq}$ inside the functional $W$ (in both the first and last arguments), and in the distribution (\ref{4.32}) as well. The condition (\ref{4.31}) applied to ${\cal F}[\Psi]$ then leads to the Euler--Lagrange equations
\begin{equation}
2 \nabla \cdot (\partial_\xi W \, \nabla \Psi)
= \partial_\tau W + \bar f.
\label{5.1}
\end{equation}
It is important here that the variation is with respect to $\Psi^V$, which ensures that there are no boundary terms. Using (\ref{4.33}), equation (\ref{4.34}) reduces to
\begin{eqnarray}
\nabla \times {\bf V}^\mathrm{eq} &=& \langle \omega({\bf r}) \rangle_W
\nonumber \\
{\bf V}^\mathrm{eq} &=& \left \langle \frac{1}{\rho_0 h({\bf r})}
\right \rangle_W \nabla \times \Psi^\mathrm{eq} + \alpha {\bf v}_\Pi.
\label{5.2}
\end{eqnarray}
Equation (\ref{5.2}) is the basic result of this section. Its solution allows one to derive the large scale mean flow encoded in $\Psi^\mathrm{eq}$ in the presence of the microscopic height and compressional fluctuations encoded in $n_{\Omega,h}({\bf r},\sigma,\lambda)$. The result therefore represents a mean field self-consistency condition, in the form of a highly nonlinear PDE, whose solution $\Psi^\mathrm{eq}({\bf r})$ also fully determines $n_{\Omega,h}$.

The solution $\Psi^\mathrm{eq}$ again depends on the Lagrange multipliers $\beta,\mu,\alpha,{\bm \gamma}$, which must be tuned to obtain prescribed values of the conserved integrals. The latter are given by (\ref{4.35}), but with all averages now with respect to $W$ at fixed $\Psi = \Psi^\mathrm{eq}$. There is one subtlety here, however. The conserved energy takes the form
\begin{eqnarray}
E &=& \frac{1}{2} \int_D d{\bf r} \left\{\left\langle
\frac{|\nabla \times \Psi^\mathrm{eq}
+ \rho_0 h \alpha {\bf v}_\Pi|^2}{\rho_0 h} \right\rangle_W \right.
\nonumber \\
&&\hskip 0.75in
\left. +\ \rho_0 g \langle \eta^2 \rangle_W
+ \frac{1}{\bar \beta} \right\}
\nonumber \\
&=& \frac{1}{2} \int_D d{\bf r} \left\{
\left\langle \frac{1}{\rho_0 h}
\right\rangle_W |\nabla \times \Psi^\mathrm{eq}|^2
+ 2 \alpha {\bf v}_\Pi \cdot \nabla \times \Psi^\mathrm{eq} \right.
\nonumber \\
&&+\ \left. \rho_0 \langle h \rangle_W \alpha^2 |{\bf v}_\Pi|^2
+ \rho_0 g \langle \eta^2 \rangle_W
+ \frac{1}{\bar \beta} \right\}.
\label{5.3}
\end{eqnarray}
The $\bar T = 1/\bar \beta$ constant term in the energy is the equipartition energy due to the quadratic fluctuations of $\Psi$ about the equilibrium value, and is produced by the Gaussian integral about saddle point in the steepest descent calculation. This term remains finite even when fluctuations are small, and represents precisely the contribution of the compressional degree of freedom $Q$ that gave rise to the $\ln(\bar \beta)$ term in (\ref{4.37}).

\subsection{Variational equations derived from the generalized Coulomb representation}
\label{sec:vareqcoulomb}

Variational equations equivalent to (\ref{5.1}) and (\ref{5.2}) can also be derived from the generalized Coulomb representation (\ref{4.37}). Since the latter is expressed entirely in terms of the original $h,\Omega$ fields, the derivation is much closer in spirit to the microcanonical approach used by RVB. Central to this approach is the local distribution function $n_{\Omega,h}({\bf r},\sigma,\lambda) = \langle \delta[\Omega({\bf r}) - \sigma] \delta[h({\bf r}) - \lambda] \rangle$ characterizing the local microscopic vorticity and height fluctuations. The grand canonical form is given in (\ref{4.32}), and the corresponding microcanonical form will be rederived here by a different route. One could in principle consider as a starting point a more fundamental three-field correlation function that includes $Q$ (see Sec.\ \ref{sec:Qdistr} below), and attempt to work directly with the original generalized Hamiltonian ${\cal K}$ defined in (\ref{4.2}). However, the divergent fluctuations in $Q \sim 1/\Delta x$ lead to the failure of the key self-averaging property used below, and hence make ${\cal K}$ a less convenient starting point. We work then with the representation (\ref{4.37}) in which $Q$ has already been integrated out.

The derivation proceeds by considering, in addition to the microscale $\Delta x$, a mesoscale $\Delta X$, both vanishing in the continuum limit, but with $\Delta X/\Delta x \to \infty$. On the scale $\Delta X$, one may define the joint probability density whose limiting form is obtained by counting the number of joint occurrences of the field levels across the $\Delta x$-cells in the given $\Delta X$-cell centered on point ${\bf r}$:
\begin{equation}
n_{\Omega,h}({\bf r},\sigma,\lambda)
= \lim_{\Delta V_g \to 0} \lim_{\Delta x \to 0}
\frac{\Delta x^2}{\Delta X^2}
\frac{\nu_{ik}}{\Delta V_g},
\label{5.4}
\end{equation}
where $i$ labels $\Delta X$-cell centers ${\bf r}_i$, $\{\sigma_k, \lambda_k \}_{k=1}^{N_g}$ is a 2D gridding of $(\Omega, h)$-space, with 2D cell volume $\Delta V_g = \Delta \Omega \Delta h$, and $\nu_{ik}$ [normalized so that $\sum_k \nu_{ik} = (\Delta X/\Delta x)^2$] counts the number of $\Delta x$-cells in $\Delta X$-cell $i$ (in the $\Delta X^2$ neighborhood of the point ${\bf r}$) with parameter value $\sigma_k,\lambda_k$ (in the $\Delta V_g$ neighborhood of $\sigma,\lambda$). The form (\ref{5.4}) ensures the normalization
\begin{equation}
\int d\sigma \int_0^\infty d\lambda \,
n_{\Omega,h}({\bf r},\sigma,\lambda) = 1
\label{5.5}
\end{equation}
and the conserved quantities are expressed in the same form (\ref{4.35}).

The phase space integral is now performed using a separation of scales: First one assigns field values for \emph{fixed} $n_{\Omega,h}$, then one integrates over all possible $n_{\Omega,h}$. The former includes all permutations of $\Delta x$-cells within a given $\Delta X$-cell (which clearly leaves $n_{\Omega,h}$ fixed, as well as all Casimirs). This sum, via the usual permutation count familiar from the lattice hard core ideal gas, produces an entropic contribution to the partition function of the form \cite{MWC1992}
\begin{eqnarray}
e^{S[n_{\Omega,h}]/\Delta x^2} &=& e^{\beta \bar T S[n_{\Omega,h}]}
\nonumber \\
S[n_{\Omega,h}] &\equiv& -\int_D d{\bf r}
\int d\sigma \int_0^\infty d\lambda
\, n_{\Omega,h}({\bf r},\sigma,\lambda)
\nonumber \\
&& \times\ \ln[R_0 n_{\Omega,h}({\bf r},\sigma,\lambda)],
\label{5.6}
\end{eqnarray}
where $R_0 = P_0/\rho_0 H_0^2$ has dimensions $[\sigma \lambda] = [\Omega h]$, and is required to make the argument of the logarithm dimensionless. This information theoretic form for $S$ is equivalent to the Sanov theorem result used by RVB.

The key assumption underlying (\ref{5.6}) is that microscale fluctuations are uncorrelated across $\Delta X$-cells: In addition to the Casimirs (which are clearly unchanged, for arbitrary shuffling of $\Omega$ values around the domain $D$), the energy and momentum should also be unchanged. Arbitrarily shuffling $h$ values, even over the entire $D$, obviously does not change the potential energy term. However, the singular fashion in which $h$ enters the Green function $G_h$ defined in (\ref{4.38}) (as well as the tensor Green function $\hat {\cal G}_h$ defined in App.\ \ref{app:KEPi}), in the form of a gradient acting on a field with $O(1)$ variations on the scale $\Delta x$, \emph{does} in fact lead to strong correlations across $\Delta X$ cells, invalidating (\ref{5.6}).

Recognizing that the result is at best approximate, we proceed now in a manner equivalent to the variational approach, by neglecting such correlations. We define the microscale averaged Green function $\overline{G_h}$ by
\begin{equation}
-\nabla \cdot \left\langle \frac{1}{\rho_0 h({\bf r})} \right\rangle_0
\nabla \overline{G_h}({\bf r},{\bf r}').
\label{5.7}
\end{equation}
Using this in place of $G_h$ one may express all quantities in terms of $n_{\Omega,h}$:
\begin{eqnarray}
E[n_{\Omega,h}] &=& \frac{1}{2} \int_D d{\bf r}
\bigg\{\frac{1}{\rho_0}
\left\langle \frac{1}{h({\bf r})} \right\rangle_0
|\langle {\bf j}({\bf r}) \rangle_0|^2
\nonumber \\
&&+\ g \rho_0 \langle [h({\bf r}) + h_b({\bf r}) - H_0]^2
\rangle_0 + \frac{1}{\bar \beta} \bigg\}
\nonumber \\
\Pi[n_{\Omega,h}] &=& \rho_0 \int d{\bf r}
[{\bf v}_\Pi({\bf r})
\cdot \langle {\bf j}({\bf r}) \rangle_0
+ \langle h({\bf r}) \rangle_0 F_\Pi({\bf r})]
\nonumber \\
\Gamma_l[n_{\Omega,h}] &=& \int_{\partial D_l}
\left\langle \frac{1}{h({\bf r})} \right\rangle_0
\langle {\bf j}({\bf r}) \rangle_0 \cdot d{\bf l}
\nonumber \\
g_\sigma[n_{\Omega,h}] &=& \int_D d{\bf r}
\int_0^\infty \lambda d\lambda \,
n_{\Omega,h}({\bf r},\sigma,\lambda).
\label{5.8}
\end{eqnarray}
Here local averages $\langle \cdot \rangle_0$ are defined in the obvious way:
\begin{equation}
\langle F[\Omega({\bf r}),h({\bf r})] \rangle_0
= \int d\sigma \int_0^\infty d\lambda F(\sigma,\lambda)
n_{\Omega,h}({\bf r},\sigma,\lambda),
\label{5.9}
\end{equation}
while the mean current density $\langle {\bf j}({\bf r}) \rangle_0$, obeying $\nabla \cdot \langle {\bf j}({\bf r}) \rangle_0$, is defined by the analog of (\ref{5.2}):
\begin{equation}
\nabla \times \left[\left\langle \frac{1}{h({\bf r})} \right\rangle_0
\langle {\bf j}({\bf r}) \rangle_0 \right]
= \langle h({\bf r}) \Omega({\bf r}) \rangle_0 + f({\bf r}),
\label{5.10}
\end{equation}
which, in addition to the circulation constraint in (\ref{5.8}), fully specifies its form. Along the same lines as (\ref{4.39}) and (\ref{4.40}), the formal solution may be expressed in terms of $\overline{G_h}$.

The total microcanonical entropy ${\cal S}$ is now given by a functional integral over all $n_{\Omega,h}$, constrained by particular values of all of the conserved quantities:
\begin{eqnarray}
e^{{\cal S}(\varepsilon,p,g)/\Delta x^2}
&=& \int D[n_{\Omega,h}] e^{S[n_{\Omega,h}]/\Delta x^2}
\delta(\varepsilon - E[n_{\Omega,h}])
\nonumber \\
&\times& \delta(p - \Pi[n_{\Omega,h}])
\prod_{l=2}^{N_D} \delta(c_l - \Gamma_l[n_{\Omega,h}])
\nonumber \\
&\times& \prod_\sigma \delta(g(\sigma) - g_\sigma[n_{\Omega,h}]),
\label{5.11}
\end{eqnarray}
The computation of ${\cal S}$ proceeds now by noting the appearance of the divergent factors $1/\Delta x^2$ in the exponentials, which produces a saddle point solution: ${\cal S}$ is the maximum of $S[n_{\Omega,h}]$ over all $n_{\Omega,h}$ obeying the constraint conditions (and, for this reason, the precise definition of the measure $\int D[n_{\Omega,h}]$ is not important here). We handle these constraints via the \emph{ordinary} use of Lagrange multipliers: rather than invoking them, via the grand canonical ensemble, at the level of the phase space integration, which introduces more stringent conditions on valid free energy minima, we use them here only to perform the constrained minimization of the functional ${\cal S}[n_{\Omega,h}]$. Thus, we introduce Lagrange multipliers $\beta = \bar \beta/\Delta x^2, \alpha, \gamma_l$, respectively, for the energy, momentum, circulation constraints, a function $\mu(\sigma)$ defining a functional
\begin{equation}
{\cal C}_\mu[n_0] = \int_D d{\bf r}
\int \mu(\sigma) d\sigma \int_0^\infty \lambda d\lambda \,
n_{\Omega,h}({\bf r},\sigma,\lambda)
\label{5.12}
\end{equation}
that is used to enforce the Casimir constraints, and an additional function $\zeta({\bf r})$ to enforce for the normalization constraint (\ref{5.5}):
\begin{equation}
{\cal N}_\zeta[n_{\Omega,h}]
= \int_D \zeta({\bf r}) d{\bf r}
\int_0^\infty d\lambda \int d\sigma \,
n_{\Omega,h}({\bf r},\sigma,\lambda).
\label{5.13}
\end{equation}
We therefore seek the minimum with respect to $n_{\Omega,h}$ of the microcanonical variational free energy
\begin{eqnarray}
{\cal F}_\mathrm{micro} &=& E[n_{\Omega,h}] - \bar T S[n_{\Omega,h}]
- \alpha \Pi[n_{\Omega,h}]
\nonumber \\
&&-\ \sum_{l=2}^{N_D} \gamma_l \Gamma_l[n_{\Omega,h}]
- {\cal C}_\mu[n_{\Omega,h}] - {\cal N}_\zeta[n_{\Omega,h}]
\nonumber \\
&&-\ 2 \bar T \int d{\bf r} \langle \ln(H_0/h({\bf r})) \rangle_0.
\label{5.14}
\end{eqnarray}
The last term accounts for the net $h^2$ factor in the phase space measure that also appears in (\ref{4.26}). The Euler-Lagrange equation, $\delta {\cal F}_\mathrm{micro}/\delta n_{\Omega,h}({\bf r},\sigma,\lambda) = 0$, produces
\begin{eqnarray}
&&\bar T \ln[(P_0/\rho_0) n_{\Omega,h}({\bf r},\sigma,\lambda)/\lambda^2]
= - \frac{\rho_0 g}{2} [\lambda + \bar h_b({\bf r}) - H_0]^2
\nonumber \\
&&\ \ \ \ \ +\ \frac{|{\bf J}({\bf r})|^2}{2\rho_0 \lambda}
+ \lambda [\mu(\sigma) - \sigma \Psi({\bf r})] - {\cal N}({\bf r}).
\label{5.15}
\end{eqnarray}
Here,
\begin{equation}
{\bf J}({\bf r}) \equiv \nabla \times \Psi({\bf r})
= \langle {\bf j}({\bf r}) \rangle_0
- \alpha \rho_0 \langle h({\bf r}) \rangle_0 {\bf v}_\Pi({\bf r})
\label{5.16}
\end{equation}
includes the momentum term, as does the shift (\ref{4.6}) to $\bar h_b$, and the circulation constraint enforces the boundary value $\Psi|_{\partial D_l} = \gamma_l$---equivalent to the first line of (\ref{4.24}). The normalization ${\cal N}({\bf r})$ combines various other constant terms with $\zeta({\bf r})$. Exponentiating this result precisely reproduces (\ref{4.32}). Inserting this result into (\ref{5.10}) produces the self-consistent variational equation for $\Psi$, equivalent to (\ref{5.2}). Inserting it into (\ref{5.8}) produces equations for the Lagrange multipliers.

\subsection{Equilibrium properties of the field $Q$}
\label{sec:Qdistr}

All of the previous results were derived by freely integrating out the compressional field $q = hQ$, resulting, via (\ref{4.19}), in $\Phi \equiv 0$ and confirming that the large scale mean flow is completely determined by the remaining field $\Psi$. The distribution $n_{\Omega,h}$, defined by (\ref{4.32}), is fundamental and allows one to compute all inputs to the variational equation (\ref{5.2}). However, it may be of interest to compute equilibrium properties of $q$ as well. As observed above, its mean $\langle q({\bf r}) \rangle = \nabla \cdot {\bf V}({\bf r})$ is trivially determined from the previously computed mean flow. More interesting are its statistical fluctuations about the mean.

To illustrate such a computation (but still within the variational approximation), we extend the two-field distribution (\ref{4.32}) to the three-field distribution function
\begin{equation}
n_0({\bf r},\sigma,\kappa,\lambda) = \langle \delta[\Omega({\bf r}) - \sigma]
\delta[\bar Q({\bf r}) - \kappa] \delta[h({\bf r}) - \lambda] \rangle,
\label{5.17}
\end{equation}
whose integral over $\kappa$ must reduce to (\ref{4.32}). Here $\bar Q = \Delta x Q$ will be seen to be the correct continuum limit scaling \cite{foot:QRVB}: the fluctuations in $Q$ are $O(1/\Delta x)$, leading to order unity fluctuations in the compressional part of the velocity ${\bf v}^C = -\nabla \phi$, and a continuous velocity potential $\phi$.

The computation again begins with the KHS-transformed free energy functional (\ref{4.17}). Integration over the field $Q$ now replaces (\ref{4.19}) by
\begin{equation}
e^{-\bar \beta \lambda \Phi({\bf r}) \kappa/\Delta x}
\prod_{{\bf r}_i \neq {\bf r}} 2\pi \delta(i\bar \beta h_i \Phi_i).
\label{5.18}
\end{equation}
where we have substituted $h({\bf r}) = \lambda$. The result of the $\Phi$ integral is again to set $\Phi_i = 0$ for all ${\bf r}_i \neq {\bf r}$, but now leaving a single nontrivial integral over $\Phi({\bf r})$. The dependence on $\Phi({\bf r})$, via the $|{\bf J}|^2/2\rho_0 h$ term, is quadratic, with $\nabla \Phi(\bar {\bf r}) = \Delta x^{-1} [\Phi(\bar {\bf r} + \Delta x {\bf \hat x}) - \Phi(\bar {\bf r}),\ \Phi(\bar {\bf r} + \Delta x {\bf \hat y}) - \Phi(\bar {\bf r})]$ nonzero only on the neighboring sites ${\bf r}$, ${\bf r}_x = {\bf r} - \Delta x {\bf \hat x}$, and ${\bf r}_y = {\bf r} - \Delta x {\bf \hat y}$. The result is the (normalized) Gaussian integral
\begin{eqnarray}
n_G({\bf r},\kappa) &=& \frac{\bar \beta \lambda}{\Delta x}
\int_C d\Phi({\bf r})
e^{-\bar \beta \lambda \Phi({\bf r})
(\kappa - \bar \kappa)/\Delta x}
\nonumber \\
&&\ \ \ \ \ \ \times\ e^{[\bar \beta \lambda \Phi({\bf r})/\Delta x]^2 \Delta \kappa^2/2}
\nonumber \\
&=& \frac{e^{-(\kappa - \bar \kappa)^2/2 \Delta \kappa^2}}
{\sqrt{2\pi \Delta \kappa^2}},
\label{5.19}
\end{eqnarray}
where we define the mean and variance
\begin{eqnarray}
\bar \kappa({\bf r},\lambda,\lambda_x,\lambda_y)
&=& \frac{1}{\rho_0 \lambda}
\left[\frac{\partial_x \Psi({\bf r})
- \partial_y \Psi({\bf r})}{\lambda} \right.
\nonumber \\
&&- \frac{\partial_x \Psi({\bf r}_y)}{\lambda_y}
+\ \left. \frac{\partial_y \Psi({\bf r}_x)}{\lambda_x}  \right]
\nonumber \\
\Delta \kappa(\lambda,\lambda_x,\lambda_y)^2
&=& \frac{1}{\bar \beta \lambda^2}
\left[\frac{2}{\lambda} + \frac{1}{\lambda_x}
+ \frac{1}{\lambda_y} \right]
\label{5.20}
\end{eqnarray}
in which $\lambda_x = h({\bf r}_x)$, $\lambda_y = h({\bf r}_y)$, and, for future reference, $\sigma_x = \Omega({\bf r}_x)$, $\sigma_y = \Omega({\bf r}_y)$ \cite{foot:dualpsi}. The result for $n_G$ is independent of $\Delta x$, as claimed.

The integral over the fields $\Omega,h$ now produce a factor $e^{\bar \beta W(\bar {\bf r})}$ [defined by (\ref{4.26}), with the same argument substitutions as in (\ref{4.28})], for every $\bar {\bf r} \notin \{{\bf r},{\bf r}_x,{\bf r}_y \}$, while the remaining integrals produce a factor
\begin{eqnarray}
n_{\bar Q}({\bf r},\kappa|\lambda)
&=& \int_0^\infty d\lambda_x \int d\sigma_x \,
n_{\Omega,h}({\bf r}_x,\sigma_x,\lambda_x)
\nonumber \\
&&\times\ \int_0^\infty d\lambda_y \int d\sigma_y \,
n_{\Omega,h}({\bf r}_y,\sigma_y,\lambda_y)
\nonumber \\
&&\ \ \ \ \ \ \times\ n_G({\bf r},\kappa|\lambda,\lambda_x,\lambda_y),
\label{5.21}
\end{eqnarray}
which differs from unity by the presence of the Gaussian factor (\ref{5.19}). One may view the result as a superposition of Gaussian densities in which the mean $\bar \kappa$ and variance $\Delta \kappa^2$ range over values weighted by the probability distribution $n_{\Omega,h}$. The normalization (\ref{5.19}) ensures that $n_{\bar Q}(\kappa)$ is a probability density for any fixed values of the other parameters.

With these inputs, the final result for $n_0$ is given by
\begin{equation}
n_0({\bf r},\sigma,\kappa,\lambda)
= n^\mathrm{eq}_{\Omega,h}({\bf r},\sigma,\lambda)
n^\mathrm{eq}_{\bar Q}({\bf r},\kappa|\lambda,\sigma)
\label{5.22}
\end{equation}
in which  ``eq'' superscript indicates that in the continuum limit one simply substitutes the variational solution $\Psi = \Psi^\mathrm{eq}$. In this same limit one may replace all appearances of $\Psi$ and its derivatives on neighboring lattice sites by their values at ${\bf r}$ wherever they appear in (\ref{5.20}) and (\ref{5.21}).

A key observation is that the fluctuation statistics predicted by $n_{\Omega,h}$ and $n_0$ are not independent. In particular, independent products of various terms for fixed height field $h$ become strongly mixed after the functional integral over $h$. This is in strong contrast to the results of RVB, in which the different choice of phase space measure does produce independent statistics \cite{foot:QRVB}. Nevertheless, despite this statistical entanglement, we have seen in Sec.\ \ref{sec:khs} that $Q$ can still be straightforwardly integrated out to produce a relatively transparent effective free energy (\ref{4.25}) for $\Omega,h$.

Note as well that the microscale Gaussian form (\ref{5.19}), and hence the precise form of $n_{\bar Q}$, is sensitive to the definition of the discrete derivative used here. One could imagine using a non-square lattice, and/or further-neighbor discrete difference forms. Given that the microscale fluctuations on the grid scale $\Delta x$ contain a finite fraction of the system energy, this sensitivity to the precise form of the grid is physically consistent. However, this sensitivity disappears upon integrating out $\kappa$. Thus, $n_{\Omega,h}$ depends only on the macroscopic flow $\Psi^\mathrm{eq}$, and produces continuum limit equilibrium forms that are insensitive to grid details. This is entirely consistent with the trivial equipartition contribution to the energy observed in (\ref{5.3}) arising from fluctuations in $Q$.

\subsection{Small height fluctuation limit}
\label{sec:smallhfluct}

In order to begin to make contact with equilibria, previously treated in the literature \cite{WP2001,CS2002}, in which surface height fluctuations were neglected, we consider here the limit in which small scale fluctuations in $h$ are assumed very small. This allows one to further reduce the problem to a simultaneous extremum problem for $\Psi$ and $h$. We continue to work within the variation approximation, though one may expect that for parameter ranges that do indeed produce small fluctuations, this approximation may often become exact (though quantifying this is beyond the scope of this paper).

Assuming that thermodynamic parameters are chosen in such a way that (\ref{4.26}) produces a very narrow distribution for $h$ about a (yet to be determined) mean, one may simplify $W$ to the form
\begin{eqnarray}
e^{\bar \beta W(\tau,h_0,\xi)}
&=& \frac{h^2}{H_0^2} e^{V(\tau,\bar \beta h)}
e^{\frac{1}{2} \bar \beta[\xi/\rho_0 h - \rho_0 g(h+h_0)^2]}
\nonumber \\
e^{V(\tau,\gamma)} &\equiv& \frac{\rho_0 H_0^3}{P_0}
\int d\sigma e^{\gamma [\mu(\sigma) - \sigma \tau]},
\label{5.23}
\end{eqnarray}
and the free energy functional (\ref{4.28}) now reduces to the form
\begin{widetext}
\begin{equation}
{\cal F}[\Psi,h] = \bar F^0(\alpha,{\bm \gamma})
- \int_D d{\bf r} \left[\frac{|\nabla \Psi|^2}{2 \rho_0 h}
- \frac{1}{2} \rho_0 g (h + \bar h_b - H_0)^2 + \bar f \Psi
+ \frac{1}{\bar \beta} V(\Psi,\bar \beta h) \right],
\label{5.24}
\end{equation}
\end{widetext}
whose minimum describes the large scale equilibrium flow, and simultaneously self-consistently determines the value of the mean surface height $h$. The Euler--Lagrange equations now produce the forms
\begin{eqnarray}
&&-\nabla \cdot \left(\frac{1}{\rho_0 h} \nabla \Psi \right) + \bar f
= -\frac{1}{\bar\beta} \partial_\tau V(\Psi,\bar \beta h)
\label{5.25} \\
&&\frac{|\nabla \Psi|^2}{2\rho_0 h^2} + \rho_0 g(h + \bar h_b - H_0)
= \partial_\gamma V(\Psi,\bar \beta h)
- \frac{2}{\bar \beta h}.
\nonumber
\end{eqnarray}
Analogous to (\ref{4.32}), the potential vorticity distribution function for given $\Psi,h$ is
\begin{eqnarray}
n_\Omega({\bf r},\sigma)
&=& \langle \delta[\Omega({\bf r}) - \sigma] \rangle
\label{5.26} \\
&=& \frac{\rho_0 H_0^3}{P_0} e^{V(\Psi,\bar\beta h)}
e^{\bar \beta h [\mu(\sigma) - \sigma \Psi)]},
\nonumber
\end{eqnarray}
from which one identifies
\begin{eqnarray}
-\frac{1}{\bar \beta} \partial_\tau V
&=& h({\bf r}) \int \sigma d\sigma \, n_\Omega({\bf r},\sigma)
\nonumber \\
&=& h({\bf r}) \langle \Omega({\bf r}) \rangle
= \langle \omega({\bf r}) \rangle + f({\bf r}).
\label{5.27}
\end{eqnarray}
Analogous to (\ref{5.2}), if we define the equilibrium flow velocity ${\bf V}$ by
\begin{equation}
{\bf V} - \alpha {\bf v}_\Pi = (\rho_0 h)^{-1} \nabla \times \Psi,
\label{5.28}
\end{equation}
the first line of (\ref{5.25}) reproduces (\ref{5.1}), while the second line produces the generalized Bernouilli equation
\begin{equation}
\frac{1}{2} \rho_0 |{\bf V} - \alpha {\bf v}_\Pi|^2 + \rho_0 g \bar \eta
= \partial_\gamma V(\Psi,\bar \beta h)
- \frac{2}{\bar \beta h}.
\label{5.29}
\end{equation}

As discussed in \cite{RVB2016}, by perturbatively treating small, but finite, fluctuations in $h$ around the mean value defined by (\ref{5.25}), the resulting theory is that of a weakly coupled system consisting of large-scale eddy motions with superimposed small scale fluctuations.

\subsubsection{Vlasov and Bernoulli conditions}
\label{subsec:vlasovbernoulli}

The forms (\ref{5.25}) appear to violate the Vlasov and Bernoulli conditions, namely that the right hand sides should depend only on the stream function $\Psi$. These conditions follow from the observations that the first line of (\ref{2.1}) and (\ref{2.7}), respectively, require that steady state flows (i.e., time-independent, in the appropriate frame of reference if momentum is conserved) obey,
\begin{eqnarray}
({\bf v}-\alpha{\bf v}_\Pi) \cdot \nabla
\left(\frac{1}{2} |{\bf v} - \alpha {\bf v}_\Pi|^2 + g \bar \eta \right) &=& 0
\nonumber \\
({\bf v}-\alpha {\bf v}_\Pi) \cdot \nabla \Omega &=& 0.
\label{5.30}
\end{eqnarray}
The steady state condition $\nabla \cdot {\bf J} = 0$ implied by the second line of (\ref{2.1}) allows one to express ${\bf J} \equiv h({\bf v}-\alpha{\bf v}_\Pi) = \nabla \times \Psi$ in terms of a current density stream function $\Psi$. Equations (\ref{5.30}) then imply that the level curves of $\Psi$, $\Omega$, and $B \equiv \frac{1}{2} |{\bf v} - \alpha {\bf v}_\Pi|^2 + g \bar \eta$ all coincide, and hence one may formally write $\Omega = f_V(\Psi)$ and $B = f_B(\Psi)$ for some fixed pair of 1D functions $f_\Omega,f_B$.

To resolve the paradox implied by the failure of the equilibrium equations to produce this functional dependence, one must understand the limits under which surface height fluctuations are small, and show that these indeed restore the Vlasov and Bernoulli conditions. We consider the cases of (1) strong gravity, $g \to \infty$, (2) low temperature $\bar \beta \to \infty$, and (3) the effects of physical processes that dissipate small scale fluctuations and hence lead to a quiescent surface.

\paragraph{Case (1):} The limit $g \to \infty$ turns out to be surprisingly subtle, and is discussed in detail in Sec.\ \ref{sec:eulercomp}. This limit indeed produces a fluctuation-free surface, $\eta \to 0$, hence $h = H_0 - \bar h_b$ independent of $\Psi$. However, due to the increased surface wave speed $c \approx \sqrt{gH_0}$, even as $g \to \infty$ one finds finite amplitude fluctuations in the compressional part of ${\bf v}$. Even though these fluctuations can still be integrated out freely [see equation (\ref{4.19})], this still leads to violations of the Vlasov condition because the advective term ${\bf v} \cdot \nabla \Omega$ in (\ref{2.7}) contains finite amplitude, correlated fluctuations in both ${\bf v}$ and $\Omega$, with the result that $\langle {\bf v} \cdot \nabla \Omega \rangle \neq {\bf V} \cdot \nabla \langle \Omega \rangle$. The same considerations apply to the Bernoulli condition, which then fails because $\nabla \cdot \langle h{\bf v} \rangle \neq \nabla \cdot (\langle h \rangle {\bf V})$.

As shown in Sec.\ \ref{sec:eulercomp}, if one imposes a strict ``rigid lid'' condition on the surface, corresponding to the Euler equation limit, the Vlasov condition is restored, but the absence of microscopic fluctuations in ${\bf v}$ and $\eta$ leads to a quantitatively different equilibrium theory. Only in the additional $\bar \beta \to \infty$ limit, discussed next, do the two theories match.

\paragraph{Case (2):} In the limit $\bar \beta \to \infty$ the term $2/\bar \beta h$ may be neglected, while a steepest descent evaluation of $V(\tau,\gamma)$ is appropriate. The latter produces
\begin{equation}
V(\tau,\gamma) \approx \gamma
\{\mu[\sigma_0(\tau)] - \tau \sigma_0(\tau) \}
\label{5.31}
\end{equation}
where $\sigma_0(\tau)$ is the solution to the stationary condition
\begin{equation}
\tau = \mu'(\sigma).
\label{5.32}
\end{equation}
This leads to
\begin{eqnarray}
&&\frac{1}{\bar \beta} V(\Psi,\bar \beta h)
\approx h \{\mu[\sigma_0(\Psi)] - \Psi \sigma_0(\Psi) \}
\nonumber \\
&&\Rightarrow\ \left\{\begin{array}{l}
\partial_\tau V/\bar \beta h = -\sigma_0(\Psi) \\
\partial_\gamma V = \mu[\sigma_0(\Psi)]
- \Psi \sigma_0(\Psi),
\end{array} \right.
\label{5.33}
\end{eqnarray}
which are indeed both functions of $\Psi$ alone. Thus, zero temperature, non-fluctuating flows indeed satisfy the requisite stream line conditions. Equation (\ref{5.32}), taking the form $\Psi = \mu'(\Omega)$, directly exhibits the Lagrange multiplier function.

\paragraph{Case (3):} This case is the most speculative, and was in fact the basis for the treatment of shallow water equilibria in Ref.\ \cite{WP2001}. There, at a critical step in the analysis, fluctuations in $h$ and $Q$ were simply assumed to have been suppressed by some set of dissipative mechanisms (e.g., viscosity, wave breaking). The resulting variational equation for $\Psi,h$ was then developed in a form similar to (\ref{5.24}) and (\ref{5.25}).

By appealing to dissipative mechanisms lying outside of the shallow water system, the theory is removed, at least temporarily, from the purely equilibrium statistical mechanics arena. The supporting notion is that in a number of physically relevant cases, a strong separation develops between the large scale eddies and the small scale wave motions, and the latter are preferentially dissipated with negligible effect on the large scale flow. The result is to remove a certain fraction of the total energy from the system, while the remainder would be proposed to lie entirely in a ``renormalized'' equilibrium flow with vanishing height fluctuations. The appropriate effective theoretical description could then be a version of case (2), in which corresponding renormalized values of the Lagrange multipliers are sought that reproduce the observed values of the conserved integrals.

An interesting consequence is that negative temperatures are no longer precluded \cite{WP2001}. Thus, $V(\tau,\gamma)$, unlike $W(\tau,h_0,\xi)$, is perfectly well defined for $\gamma < 0$, and so extrema of (\ref{5.24}) may be sought for both positive and negative $\bar \beta$ (in particular, for both $\bar \beta \to \pm \infty$). In principle, negative temperature equilibria are unstable to leakage of energy into (positive temperature) wave motions, but the physical coupling of large scale flows to small scale wave generation is extremely small, and it makes sense to develop a theory along these lines that neglects such effects. The key observation here is that compact eddy structures, such as Jupiter's Great Red Spot, having vorticity maxima confined away from the system boundaries, can only be interpreted as negative temperature states \cite{MWC1992}. Such structures therefore lie outside the strict shallow water theory presented here, and nonequilibrium dissipation arguments \emph{must} therefore be invoked in order to make contact with the effective equilibrium descriptions ubiquitous in the literature \cite{BV2012}.

We note finally that there is no reason for the more general result (\ref{5.1})---or, for that matter, the fully fluctuating result (\ref{4.34})---to satisfy these conditions because the microscale flows are not steady state. The conditions need only be restored when such fluctuations are assumed to be absent. An interesting point is that RVB found that, even in their general theory, both conditions to be satisfied, and cite this as a supporting feature \cite{RVB2016}. Their result occurs because, in contrast to (\ref{5.2}), their version of the phase space measure produces independent microscale fluctuations of $h,Q,\Omega$. This leads in particular to $\langle {\bf v} \cdot \nabla Q \rangle = {\bf V} \cdot \nabla \langle Q \rangle$ and $\langle \nabla \cdot (h{\bf v}) = \nabla \cdot (\langle h \rangle {\bf V})$, and this is then reflected in the desired $\Psi$-dependence of the equilibrium equations. The feature therefore is a direct consequence of the inconsistency of their measure choice with the Liouville theorem (see Apps.\ \ref{app:liouville} and \ref{app:liouvilleinequiv}), and we argue therefore that it should not (in absence of much deeper arguments) be considered as supporting the validity of the approach.

\section{Comparison with Euler equilibria}
\label{sec:eulercomp}

In this section, shallow water equilibria will be compared to those of the Euler equation, including variable bottom topography $h_b({\bf r})$, but now with a fixed rigid-lid surface. It will be shown that the latter leads to an equilibrium phase space measure with non-uniform gridding of the domain $D$, determined by $h_b$---consistent in this case with the choice made by RVB. This is significantly different from the limit $g \to \infty$ in the shallow water results of Sec.\ \ref{sec:statmech}, which continues to require uniform gridding. The paradox is resolved by showing that the equilibria are in fact expected to be physically \emph{different}, with microscale height fluctuation effects present even in the limit $g \to \infty$. These results serve again to highlight the inconsistency of the RVB nonuniform grid choice with that implied by the shallow water Liouville equation.

The Euler equation, including variable bottom topography, is described by
\begin{eqnarray}
\partial_t {\bf v} + ({\bf v} \cdot \nabla) {\bf v}
+ f {\bf \hat z} \times {\bf v} &=& -\frac{1}{\rho_0} \nabla p
\nonumber \\
\nabla \cdot (h {\bf v}) &=& 0,
\label{6.1}
\end{eqnarray}
and is equivalent to the shallow water equations (\ref{2.1}) but with $h({\bf r}) = H_0 - h_b({\bf r})$ now a fixed function, and the pressure $p$ enforcing the incompressibility condition (and an equation for which is obtained by multiplying both sides of the first line by $h$ and taking the divergence). The potential vorticity is still given by (\ref{2.6}), and continues to be advectively conserved [equation (\ref{2.7})].

The incompressibility condition implies that
\begin{equation}
{\bf j} = \rho_0 h {\bf v} = \nabla \times \psi
\label{6.2}
\end{equation}
is purely transverse. The velocity and potential vorticity
\begin{eqnarray}
{\bf v} &=& \frac{1}{\rho_0 h} \nabla \times \psi
\nonumber \\
h\Omega - f &=& -\nabla \cdot \left(\frac{1}{\rho_0 h} \nabla \psi \right)
\label{6.3}
\end{eqnarray}
are completely determined in terms of the single scalar function $\psi$. The equation of motion (\ref{2.7}) therefore fully describes the Euler dynamics. Statistical equilibria, obeying the Vlasov condition
\begin{equation}
{\bf v} \cdot \nabla \Omega = 0,
\label{6.4}
\end{equation}
are then formulated entirely in terms of $\Omega$ as well. For simplicity, we will consider only a simply connected domain $D$ with Dirichlet boundary conditions on $\psi$. Defining the (symmetric) scalar Green function $G_h$ by
\begin{equation}
-\nabla \cdot \frac{1}{\rho_0 h} \nabla G_h({\bf r},{\bf r}')
= \delta({\bf r}-{\bf r}')
\label{6.5}
\end{equation}
with Dirichlet boundary conditions [identical to (\ref{4.38}), but now with deterministic $h$], one obtains the relation
\begin{equation}
\psi({\bf r}) = \int_D d{\bf r} G_h({\bf r},{\bf r}')
(h\Omega - f)({\bf r}').
\label{6.6}
\end{equation}

\subsection{Liouville theorem and equilibrium measures}
\label{sec:eulerliouville}

The Liouville theorem follows from the equation of motion written in the conserved form
\begin{equation}
h \dot \Omega = -\nabla \cdot [h \Omega {\bf v}],
\label{6.7}
\end{equation}
which leads to \cite{foot:liouville}
\begin{equation}
h({\bf r}) \frac{\delta \dot \Omega({\bf r})}{\delta \Omega({\bf r})}
= - \nabla \cdot \left\{h({\bf r})
\frac{\delta [\Omega({\bf r}) {\bf v}({\bf r})]}{\delta \Omega({\bf r})} \right\}.
\label{6.8}
\end{equation}
From the boundary conditions on $\partial D$, it follows that
\begin{equation}
\int_D d{\bf r} h({\bf r})
\frac{\delta \dot \Omega({\bf r})}{\delta \Omega({\bf r})} = 0.
\label{6.9}
\end{equation}
In order to express this in the standard form of a divergence free condition on the phase space flows, we define a mapping ${\bf r}({\bf a}): D \to D$ (clearly not unique, but still fixed by the bottom topography) with Jacobian
\begin{equation}
J({\bf a}) \equiv \frac{\partial {\bf r}}{\partial {\bf a}}
= \frac{H_0}{h[{\bf r}({\bf a})]}.
\label{6.10}
\end{equation}
Thus, ${\bf r}({\bf a})$ maps a fluid with uniform height $H_0$ to one with variable, but time-independent, height $h$. Relabeling $\Omega({\bf a}) \equiv \Omega[{\bf r}({\bf a})]$, (\ref{6.9}) may be written in the form
\begin{equation}
\int_D d{\bf a}
\frac{\delta \dot \Omega({\bf a})}{\delta \Omega({\bf a})} = 0.
\label{6.11}
\end{equation}

It follows immediately from (\ref{6.11}) that equilibrium statistical measures $\rho(E,P,{\cal C})$ are, as usual, functions only of the conserved integrals, and phase space averages may be defined through the continuum limit
\begin{equation}
\rho[\Omega] D[\Omega]
= \lim_{\Delta V \to 0} \rho[\Omega] \prod_i d\Omega_i,
\label{6.12}
\end{equation}
in which $\Omega_i = \Omega({\bf a}_i)$ and $\{{\bf a}_i \}$ represents a \emph{uniform} gridding of $D$, with fixed physical fluid element volume $\Delta V = H_0 \Delta A =  h_i \Delta A_i$, where $\Delta A_i$ is the image of cell $i$ under the mapping ${\bf r}({\bf a})$.

\subsection{Statistical mechanics}
\label{sec:eulerstatmech}

The grand canonical statistical measure is given by
\begin{eqnarray}
\rho &=& \frac{1}{Z} e^{-\beta {\cal K}[\Omega]}
\nonumber \\
{\cal K}[\Omega] &=& E[\Omega] - C_\mu[\Omega]
\label{6.13}
\end{eqnarray}
with energy and Casimir functionals
\begin{eqnarray}
E[\Omega] &=& \frac{\rho_0}{2} \int_D d{\bf r} \int_D d{\bf r}'
(h \Omega - f)({\bf r})
\nonumber \\
&&\ \ \ \ \ \ \ \ \times G_h({\bf r},{\bf r}') (h \Omega - f)({\bf r}')
\nonumber \\
C_\mu[\Omega] &=& \int_D d{\bf r} h({\bf r}) \mu[\Omega({\bf r})].
\label{6.14}
\end{eqnarray}
Given the simply connected domain, momentum conservation is not considered here.

In discrete form, one obtains
\begin{eqnarray}
E &=& \int_D d{\bf a} \int_D d{\bf a}' (\Omega - f/h)({\bf a})
\nonumber \\
&&\ \ \ \ \ \times G_h[{\bf r}({\bf a}),{\bf r}({\bf a}')]
(\Omega - f/h)({\bf a}')
\nonumber \\
&=& \lim_{\Delta V \to 0} \frac{\rho_0}{2} \Delta V^2 \sum_{i,j}
(\Omega - f/h)_i G_{h,ij} (\Omega - f/h)_j
\nonumber \\
C_\mu &=& H_0 \int_D d{\bf a} f[\Omega({\bf a})]
= \lim_{\Delta V \to 0} \Delta V \sum_i \mu[\Omega({\bf a}_i)].
\nonumber \\
\label{6.15}
\end{eqnarray}
The KHS transformation, acting to decouple the energy, produces the partition function
\begin{eqnarray}
Z &=& \prod_i \frac{P_0}{\rho_0 H_0^3} \int d\Omega_i e^{-\beta {\cal K}[\Omega]}
\nonumber \\
&=& \frac{1}{{\cal N}_h} \prod_i P_0 H_0 \int d\Psi_i e^{-\beta {\cal F}[\Psi]}
\label{6.16}
\end{eqnarray}
with (continuum limit) free energy functional
\begin{equation}
{\cal F}[\Psi] = -\int d{\bf r}
\left[\frac{|\nabla \Psi|^2}{2 \rho_0 h}
+ f \Psi + h W(\Psi) \right].
\label{6.17}
\end{equation}
The normalization ${\cal N}_h$ is the determinant of the quadratic form that defines $E$ in (\ref{6.15}), and is a nontrivial functional of $h$. This is contrast to shallow water KHS result (\ref{4.10}) where the normalization takes the form of trivial product factors. However, it is a fixed constant, and hence does not contribute to equilibrium averages.

The Lagrange multiplier function $\mu$ is now subsumed into the function $W(\tau)$ defined by
\begin{equation}
e^{\bar \beta_E W(\tau)}
= \int d\sigma e^{\bar \beta_E [\mu(\sigma) - \sigma \tau]}
\label{6.18}
\end{equation}
with renormalized temperature variable
\begin{equation}
\bar \beta_E = \beta \Delta V,\ \bar T_E = T/\Delta V
\label{6.19}
\end{equation}
remaining finite in the continuum limit $\Delta V \to 0$. Both positive and negative temperatures are allowed here since convergence of the integral is in general controlled by $\mu(\sigma)$. In this limit, one has $|\beta| \to \infty$ and the variational condition where one seeks the minimum of ${\cal F}$ emerges. In this case, since the compressional degree of freedom has been suppressed at the outset and, correspondingly, the height field is fixed, $\Psi$ is continuously differentiable, and both $\Psi$ and $\nabla \Psi$ are non-fluctuating in the continuum limit (while, of course, $\nabla^2 \Psi$ has finite fluctuations). The variational approximation is therefore exact in this case, and one obtains the Euler-Lagrange equation
\begin{equation}
\omega_0 \equiv -\nabla \cdot \left(\frac{1}{\rho_0 h} \nabla \Psi_0 \right)
= -h W'(\Psi_0) - f.
\label{6.20}
\end{equation}
The equilibrium potential vorticity obeys
\begin{equation}
\Omega_0 = \frac{\omega_0 + f}{h} = -W'(\Psi_0),
\label{6.21}
\end{equation}
which ensures that the Vlasov condition (\ref{6.4}) is satisfied (i.e., $\nabla \Psi_0$ and $\nabla Q_0$ are everywhere colinear, and hence $\Psi_0$ and $\Omega_0$ share stream lines).

\subsection{Comments}
\label{sec:comments}

Unlike for the shallow water equations, in which the microscale (especially surface height) fluctuations contribute in a highly nontrivial way, even in the $g \to \infty$ limit, to the variational free energy (\ref{4.28}) for the large-scale vortical stream function, the rigid lid boundary conditions here suppress these entirely, and the Vlasov condition is satisfied explicitly. The key enabling result is that the velocity ${\bf v} = {\bf v}_0$ is purely large scale.

From a mathematical point of view, the Vlasov result \emph{requires} that the function $W$ have spatial dependence through $\Psi$ alone---all dependence on $h$ escapes only to the overall multiplier of $W(\Psi)$ in (\ref{6.17}). This happens only because the Liouville equation that determines the phase space measure produces, in contrast to the shallow water case, a \emph{nonuniform} real space mesh. Intuitively, the rigid lid condition places corresponding rigid conditions on the Eulerian phase space fluid parcel distribution $\nu({\bf r},{\bf p})$ defined in App.\ \ref{app:liouville}. Specifically, the moments defined in (\ref{A18}) are restricted by $h = H_0 - h_b$ and $\nabla \cdot {\bf j} = 0$. The resulting reformulation (\ref{6.12}) entirely in terms of $\Omega$, which enforces these conditions automatically, then also induces the nonuniform mesh.

Comparing to the $g \to \infty$ limit of the shallow water equations, one observes that the function $W$ in (\ref{4.26}) or $V$ in (\ref{5.23}) continues to depend nontrivially on $h$, even though the condition $\eta = h + h_b - H_0 \to 0$ is indeed enforced, in the small fluctuation limit, through the $g (h+h_0)^2$ term in (\ref{4.26}) or $g \bar \eta^2$ term in (\ref{5.24}). The resolution of this paradox is that although $\eta/H_0 = O(1/\sqrt{\bar \beta \rho_0 g H_0^2})$ is very small, the compressional part of the velocity ${\bf v}_L = O(c \eta/H_0) = O(1/\sqrt{\bar \beta \rho_0 H_0})$ remains finite because the wave speed $c = \sqrt{g H_0}$ (along with the microscopic frequency $\partial_t h/H_0$) diverges. Thus, the Vlasov combination ${\bf v} \cdot \nabla \Omega$ has finite amplitude (correlated) microscopic fluctuations in \emph{both} ${\bf v}$ and $\Omega$, and the equilibrium average $\langle {\bf v} \cdot \nabla \Omega \rangle \neq {\bf v}_0 \cdot \nabla \Omega_0$ fails to factorize [except in the additional low temperature limit $\bar \beta \to \infty$ described by (\ref{5.31})--(\ref{5.33})]. This explains the violation of the Vlasov condition implied by the dependence on $h$ of the right hand side of the first line of (\ref{5.25}).

As a final comment, we note that the heuristic suppression of small-scale wave fluctuations considered in Ref.\ \cite{WP2001} also led to a theory with a nonuniform $h$-dependent mesh. However, in this case $h$ was not fixed \emph{a priori}, but determined, along with $\Psi$, through the free energy minimization (which, as we have seen, implicitly assumes that all of the energy is in the large scale flow, and hence provides the mathematical mechanism for suppressing small-scale waves). Self-consistently, the resulting hydrostatically balanced flows satisfied both the Vlasov and Bernoulli conditions.

\section{Concluding remarks}
\label{sec:conclude}

An important distinction between RVB and the present approach is the use here of the grand canonical ensemble, and of the KHS transformation (Sec.\ \ref{sec:khs}), as key tools for deriving useful reduced forms for the effective free energy functional from the generalized Hamiltonian (\ref{4.2}). By expressing the generalized Hamiltonian in the purely local form (\ref{4.17}), the method has the advantage of providing a mathematically complete and efficient procedure for deriving the intermediate reduced form (\ref{4.25}) (integrating out $Q,\Phi$), following with either the fully reduced elastic membrane form (\ref{4.28}) (integrating out $h,\Omega$, leaving only $\Psi$), or the dual generalized Coulomb form (\ref{4.37}) (integrating out $\Psi$, leaving $\Omega,h$). Most importantly, it transparently exhibits the strong fluctuations and long-range correlations that survive the continuum limit. The formulation adopted by RVB misses both of these effects because the mean field approximation is implicit in their approach to separating the fields into large scale and small scale components.

The discussion in App.\ \ref{app:liouvilleinequiv} on the connection between the Liouville theorem and equilibrium measures is based on a very general formulation (\ref{B2}) or (\ref{B10}) of the Liouville theorem, and does not require an appeal to an underlying Hamiltonian structure. The latter is used as part of the specific derivation in App.\ \ref{app:liouville}, but the conclusions follow much more generally. In particular, the theory leads quite generally to the construction of a phase space measure through a limiting procedure completely consistent with standard uniform area gridding of the field index ${\bf r}$, which is also fully consistent with many previous statistical mechanics applications in quantum and classical field theory.

RVB instead replace the uniform grid by a highly nonuniform ``Lagrangian'' grid, that is moreover \emph{dynamically adjusted} according to the fluid height field, which is itself one of the phase space variables being integrated over. Given that $h$ has strong variations on the grid scale, this is a rather singular adjustment, and is very unlike, for example, the smooth change of variable adopted in Ref.\ \cite{WP2001} after the microscale height fluctuations were assumed to have been dissipated, or the time-independent change of variable(\ref{6.10}) in the Euler case (with degree of smoothness governed by the bottom topography $h_b$). We have seen that the RVB choice corresponds to a very different form of the Liouville theorem---equivalent to a nontrivial density $w({\bf r})$ in (\ref{B17}) that also includes strong variations on the grid scale.

The equilibrium theory resulting from the two choices are quantitatively different, so this is not an instance of mathematical convenience to obtain an equivalent continuum limit. In particular, we have emphasized that the shallow water equilibrium states \emph{are not expected to be stationary, time independent solutions to the fluid equations.} Unlike the pure 2D Euler case (discussed in detail in Sec.\ \ref{sec:eulercomp}), we have seen that the macroscopic flows are strongly dynamic, with finite energy, finite amplitude, high frequency height fluctuations (resulting from the undissipated forward cascade of wave energy). They are found to be stationary in Ref.\ \cite{RVB2016} only because the Lagrangian gridding leads to a product measure in which the basic fields have independent statistics, leading to exact factorization of key averages. In the present theory, the height field $h$ is not independent of ${\bf v}$, and this leads to the expected nonstationary equilibrium averages. As discussed in Sec.\ \ref{sec:eulercomp}, this is also what leads to the inequivalence of the rigid lid Euler flow and shallow water $g \to \infty$ limit.

This paper has concentrated on deriving general statistical models and exploring some of their key general features. Detailed studies of equilibrium solutions for specific, physically motivated choices of model parameters remains to be addressed in future work.
The effects of fluctuations, and predicting the effects of various dissipation mechanisms in producing the ultimate \emph{quiescent} equilibria \cite{WP2001}, deserve special focus. RVB have already made some explorations along these lines within the variational theory.
Significant insight can be gained by restricting the problem to a finite number of degrees of freedom. For example, the choice $\mu(\sigma) = -\frac{1}{2} \mu_0 \sigma^2$ reduces (\ref{4.26}) to a Gaussian integral in the variable $\sigma$, and corresponds to a version of the Energy--Enstrophy theory \cite{K1975}. Perhaps more interesting are the finite-level systems $e^{\bar \beta \mu(\sigma)} = \sum_{n=1}^{N_\sigma} e^{\bar \beta \mu_n} \delta(\sigma - \sigma_n)$ \cite{MWC1992,BV2012} in which the potential vorticity is permitted to take only a discrete set of values, with relative populations controlled by the corresponding discrete set of chemical potentials $\mu_n$. Even the cases $N_\sigma = 2,3$ generate an interesting variety of equilibria as the temperature and other parameters are varied.

Most previous investigations have focused on mean field equilibria, especially those of the Euler and quasigeostrophic equations for which they are exact. A very interesting feature is the set of transitions between equilibrium states that can occur as a function of the thermodynamic parameters. An important example is when a translation or rotational symmetry is broken: with increasing energy, an instability can occur in which an annular or linear jet transitions to a more compact vortex structure. Within the variational approximation, such transitions are simple bifurcations. In the presence of strong fluctuations the character of the transition remains an open question. A possibility is that it elevates to a true critical phenomenon with nontrivial critical exponents \cite{S1971}. Phase transitions in the context of elastic membranes include roughening of  crystalline solid facets \cite{CL1995}. Here there is competition between a periodic confining potential which prefers a flat interface, and entropic fluctuations which prefer a rough surface with the logarithmic correlations alluded to in Sec.\ \ref{subsec:nonlinmembrane}. In the present case the analogue of a periodic crystalline potential is absent, and the membrane is always in the rough regime. Instead, there is a large-scale conformational change of the membrane, more akin perhaps to shape changes in biological membrane systems \cite{PKTH2013}.

\appendix

\section{Fluid system Liouville theorem and phase space measure}
\label{app:liouville}

In this Appendix a very general 2D fluid system Liouville theorem is derived, applicable to a much more general class of equations than just the shallow water system. The derivation is based on a Lagrangian coordinate description, in which standard Hamiltonian position and conjugate momentum coordinates may be transparently derived and applied \cite{WP2001}. A transformation to Eulerian coordinates is made at the end to demonstrate equivalence for the special case, specific to the shallow water equations, derived in Ref.\ \cite{RVB2016}. The equivalence lends insight to the the nature of the microscale fluctuations being considered.

\subsection{Lagrangian coordinate Hamiltonian formulation}
\label{app:lagham}

In the presence of both a Coriolis force $f({\bf r})$ and bottom topography $h_b({\bf r})$, the Lagrangian coordinate Hamiltonian takes the form
\begin{eqnarray}
{\cal H} &=& \int_D d^2a \left(\frac{|{\bf p}({\bf a})
- {\bf A}[{\bf r}({\bf a})]|^2}{2\rho_0 H_0} \right.
\nonumber \\
&&+\ \left. \frac{1}{2} \rho_0 H_0 g
\{h[{\bf r}({\bf a})] + 2 h_b[{\bf r}(a)] \}
\right)
\label{A1}
\end{eqnarray}
where, incorporating a minor rescaling from that appearing in (\ref{2.16}) and (\ref{2.19}), the vector potential represents the Coriolis parameter via
\begin{equation}
\rho_0 H_0 f({\bf r}) = \nabla \times {\bf A}({\bf r})
= \partial_x A_y - \partial_y A_x,
\label{A2}
\end{equation}
and has the physical interpretation of the steady velocity field that produces the background coordinate system rotation. Units have been chosen so that ${\bf p}$ is an areal momentum density, i.e., it has the same dimensions as ${\bf j}$.

The Lagrangian coordinate ${\bf r}({\bf a},t)$ and conjugate momentum ${\bf p}({\bf a},t)$ represents a fluid parcel of fixed volume $H_0 d^2a$. For each $t$, ${\bf a} \to {\bf r}({\bf a},t)$ represents a mapping of the domain $D$ into itself. Unlike for the case of the Euler equation, this mapping is not in general area preserving. In fact, the height field is defined by
\begin{equation}
\frac{H_0}{h[{\bf r}({\bf a})]} = J({\bf a})
\label{A3}
\end{equation}
where
\begin{equation}
J({\bf a}) = \det\left(\frac{\partial {\bf r}}{\partial {\bf a}}\right)
= (\partial_1 r_1)(\partial_2 r_2) - (\partial_2 r_1)(\partial_1 r_2)
\label{A4}
\end{equation}
is the Jacobian of the transformation. Thus, in a slight abuse of notation, $h[{\bf r}({\bf a})]$ is actually a nontrivial functional of ${\bf r}({\bf a})$. The corresponding term in ${\cal H}$ represents the potential energy of a particle at height $h/2$ above the bottom, equivalent to that of a fluid parcel of thickness $h$.

Hamilton's equations of motion then yield
\begin{eqnarray}
\dot {\bf r}({\bf a}) &=& \frac{\delta {\cal H}}{\delta {\bf p}({\bf a})}
= \frac{{\bf p}({\bf a}) - {\bf A}[{\bf r}({\bf a})]}{\rho_0 H_0}
\nonumber \\
\dot {\bf p}({\bf a}) &=& -\frac{\delta {\cal H}}{\delta {\bf r}({\bf a})}
\nonumber \\
&=& \frac{1}{\rho_0 H_0} (\nabla {\bf A})[{\bf r}({\bf a})]
\cdot \{{\bf p}({\bf a}) - {\bf A}[{\bf r}({\bf a})]\}
\nonumber \\
&&-\ \rho_0 H_0 g \nabla \{h[{\bf r}({\bf a})] + h_b[{\bf r}({\bf a})] \},
\nonumber \\
\label{A5}
\end{eqnarray}
where the gradient of $h$ is defined through the change of variable
\begin{equation}
\nabla h[{\bf r}({\bf a})]
= \left(\frac{\partial {\bf r}}{\partial {\bf a}}\right)^{-1}
\nabla_a h[{\bf r}({\bf a})].
\label{A6}
\end{equation}
Newton's equation of motion are then obtained in the form
\begin{eqnarray}
\ddot {\bf r}({\bf a}) &=& \frac{1}{\rho_0 H_0}
\{\dot {\bf p}({\bf a})
- [\dot {\bf r}({\bf a}) \cdot \nabla] {\bf A}[{\bf r}({\bf a})]\}
\label{A7} \\
&=& -f[{\bf r}({\bf a})] {\bf \hat z} \times \dot {\bf r}({\bf a})
- g \nabla \{h[{\bf r}({\bf a})] + h_b[{\bf r}({\bf a})] \},
\nonumber
\end{eqnarray}
where the Coriolis term has been produced by the antisymmetric combination
\begin{eqnarray}
&&(\nabla A) \cdot \dot {\bf r}
- \dot {\bf r} \cdot(\nabla A)
= [\nabla A - (\nabla A)^T] \cdot \dot {\bf r}
\nonumber \\
&&\ \ \ \ \ =\ \left(\begin{array}{cc}
0 & \partial_y A_x - \partial_x A_y \\
\partial_x A_y - \partial_y A_x & 0
\end{array} \right) \dot {\bf r}
\nonumber \\
&&\ \ \ \ \ =\ -\rho_0 H_0 f {\bf \hat z} \times \dot {\bf r}.
\label{A8}
\end{eqnarray}

Equation (\ref{A7}) is the Lagrangian equivalent of the first line of (\ref{2.1}), where one identifies ${\bf v}[{\bf r}({\bf a},t),t] = \dot {\bf r}({\bf a},t)$ and $\frac{d}{dt} = \partial_t + {\bf v} \cdot \nabla$. The second line of (\ref{2.1}) follows from the equation of motion for $h[{\bf r}({\bf a},t)]$:
\begin{eqnarray}
\frac{d}{dt} h[{\bf r}({\bf a},t)]
&=& - h[{\bf r}({\bf a},t)] \mathrm{tr}
\left[\left(\frac{\partial {\bf r}}{\partial {\bf a}} \right)^{-1}
\nabla_a \dot {\bf r}({\bf a},t) \right]
\nonumber \\
&=& - h[{\bf r}({\bf a},t)] \nabla \cdot \dot {\bf r}({\bf a},t).
\label{A9}
\end{eqnarray}

\subsection{Lagrangian coordinate Liouville theorem}
\label{app:liouvillelagrange}

For any Hamiltonian system, the phase space invariant measures take the form
\begin{equation}
d\Gamma = \rho({\cal H},\{{\cal C}_i\}) D[{\bf p}] D[{\bf r}]
\label{A10}
\end{equation}
where the phase space density $\rho$ is an arbitrary function of all of the conserved integrals of the motion (the proof is outlined in a more general context in App.\ \ref{app:liouvilleinequiv}). Here $\{{\cal C}_i\}$ represents the collection of all conserved quantities, besides ${\cal H}$, including total momentum or angular momentum, depending on the domain symmetries, as well as the Casimirs (\ref{2.10}), and the circulations (\ref{2.13}). The choice of $\rho$ determines the ensemble. The functional integration measure, consisting of an independent product over all coordinates and momenta, may be defined by the continuum limit
\begin{equation}
D[{\bf r}] D[{\bf p}] = \lim_{b \to 0} \frac{1}{N_b!} \prod_{j=1}^{N_b}
\frac{d{\bf p}({\bf a}_j) \, d{\bf r}({\bf a}_j)}{(P_0 b)^2},
\label{A11}
\end{equation}
in which the domain $D$ is approximated by a square mesh of $N_b = A_D/b^2$ fluid parcels of \emph{equal area} $b^2$ (and equal volume $H_0 b^2$). The $1/N_b!$ factor accounts for the arbitrary parcel relabeling symmetry. The factor $1/(P_0 b)^2$ is included to obtain a dimensionless partition function, and a properly normalized free energy, with $P_0$ an arbitrary constant with the same units as ${\bf p}$ (e.g., $P_0 = \rho_0 H_0 V_0$, where $V_0$ is a characteristic fluid velocity).

\subsection{Conversion to Eulerian coordinates}
\label{app:eulerconv}

The alternative Eulerian formulation is obtained by recognizing that the coordinate measure $D[{\bf r}]D[{\bf p}]$ corresponds to a physical coordinate--momentum space \emph{Poisson process}, in which each fluid parcel is placed independently, with uniform probability, in the domain ${\cal D} = D \times \mathbb{R}^2$, i.e., at a particular point ${\bf r}({\bf a})$ in $D$ with a particular momentum ${\bf p}({\bf a})$. In order to obtain a sensible limit, we can restrict ${\bf p}$ to a compact domain $D_P \subset \mathbb{R}^2$, with finite area $A_P$, and take the limit $D_P \uparrow \mathbb{R}^2$ in the end.

An equivalent statistical description divides ${\cal D}$ into an arbitrary fixed mesh, and counts the number of parcels $n_{lm} = n({\bf r}_l,{\bf p}_m)$ in each cell of phase space volume $\Delta V = \Delta p^2 \Delta x^2$. The limit $b \to 0$ will be taken first first, at fixed $\Delta x, \Delta p$, so that $n_{lm} \to \infty$. To obtain a sensible limit, one defines the continuous variables,
\begin{equation}
\nu_{lm} = \frac{P_0^2 b^2}{\Delta x^2 \Delta p^2} n_{lm}
\label{A12}
\end{equation}
The partition function is then obtained by freely and independently integrating over each $\nu_{lm}$. The constraint $\sum_{l,m} n_{lm} = N_b$ leads to
\begin{equation}
\sum_{l,m} \nu_{lm} = \frac{P_0^2 A_D}{\Delta p^2 \Delta x^2}.
\label{A13}
\end{equation}
With this normalization, the continuum limit $\Delta x, \Delta p \to 0$ produces $\nu_{lm} \to \nu({\bf p},{\bf r})$, with
\begin{equation}
\sum_{l,m} \nu_{lm} \Delta x^2 \Delta p^2
\to \int d{\bf p} \int_D d{\bf r} \,
\nu({\bf r},{\bf p}) = P_0^2 A_D.
\label{A14}
\end{equation}
This continuum notation is intended here only as a heuristic, because $\nu_{lm}$ fluctuates wildly from cell to cell. However, the notational intent is clear, and the more rigorous mathematical statement resides only in the underlying finite dimensional calculations that are then used to derive well defined, smoothly varying, limiting forms for thermodynamic averages.

The corresponding Eulerian phase space measure is now given by an integration over the phase space defined by all functions $\nu$:
\begin{equation}
d\Gamma  = \rho({\cal H},\{{\cal C}_i\}) D[\nu]
\label{A15}
\end{equation}
defined by the continuum limit
\begin{equation}
D[\nu] = \lim_{\Delta V \to 0} \prod_{l,m} d\nu_{lm}.
\label{A16}
\end{equation}
The usual Eulerian hydrodynamic fields are obtained from the moments
\begin{eqnarray}
h_l &=& \frac{H_0 b^2}{\Delta x^2} \sum_m n_{lm}
= \frac{H_0 \Delta p^2}{P_0^2} \sum_m \nu_{lm}
\nonumber \\
{\bf j}_l &=& {b^2}{\Delta x^2} \sum_m n_{lm} {\bf p}_m
= \frac{\Delta p^2}{P_0^2} \sum_m \nu_{lm} {\bf p}_m.
\label{A17}
\end{eqnarray}
In continuum notation,
\begin{eqnarray}
h({\bf r}) &=& H_0 \int \frac{d{\bf p}}{P_0^2} \nu({\bf r},{\bf p})
\nonumber \\
{\bf j}({\bf r}) &=& \int \frac{d{\bf p}}{P_0^2} {\bf p}\,
\nu({\bf r},{\bf p})
\label{A18}
\end{eqnarray}
The velocity is defined as usual by ${\bf j} = \rho_0 h {\bf v}$.

\subsection{Reduced moment description}
\label{app:reducedmoment}

In principle, the phase space measure (\ref{A15}) is applicable to any Hamiltonian and conserved integrals constructed as arbitrary functionals of $\nu$. However, if the phase space integrand depends only on the two moments (\ref{A18}), one may reduce (\ref{A15}) by integrating out all other degrees of freedom. This is accomplished formally by representing
\begin{eqnarray}
\nu({\bf r},{\bf p}) &=& \sum_m M_m({\bf p}) \nu_m({\bf r})
\nonumber \\
\nu_m({\bf r}) &\equiv & \int d{\bf p} M_m({\bf p}) \nu({\bf r},{\bf p})
\label{A19}
\end{eqnarray}
as an expansion in a complete set of moment functions $M_n({\bf p})$ (e.g., Legendre polynomials covering the domain $D_P$), constrained by the choice $M_0 = 1$, $M_1 = p_x$, $M_2 = p_y$. One may then write
\begin{equation}
d\Gamma  = \rho({\cal H},\{{\cal C}_i\}) \prod_n D[\nu_n].
\label{A20}
\end{equation}
Since $\rho$ depends only on $\nu_0 = h$ and $(\nu_1,\nu_2) = {\bf j}$, one may freely integrate out all higher $\nu_n$, $n > 2$, to obtain the ``reduced'' Eulerian phase space measure
\begin{equation}
d\Gamma_E  = \rho({\cal H},\{{\cal C}_i\}) D[h] D[{\bf j}]
\label{A21}
\end{equation}
defined by the continuum limit
\begin{equation}
D[h] D[{\bf j}] = \lim_{\Delta x \to 0}
\prod_l \frac{dh_l \, d{\bf j}_l}{H_0 P_0^2}.
\label{A22}
\end{equation}
The factor $1/H_0 P_0^2$ is again included for dimensional purposes. Unlike (\ref{A11}), which includes the parcel area $b^2 \to 0$, the microgrid area $\Delta x^2$ does not appear in (\ref{A22}) because there is no $1/N_E!$ combinatorial factor in this representation. Note that the constraint of fixed fluid volume $H_0 A_D = \sum_l h_l \Delta x^2$ is already accounted for in $\rho$ through the Casimir (\ref{2.9}) with $w \equiv 1$.

The form (\ref{A22}) may also be derived directly from the Eulerian fluid equations (\ref{2.1}) \cite{RVB2016}. However, the derivation here exhibits the possibility of a much more general class of Hamiltonians and conserved integrals that could depend on the full $\nu({\bf r},{\bf p})$, not just a few of its moments.

\subsection{Potential vorticity description}
\label{app:potvort}

Significant work was done in Sec.\ \ref{sec:canonfields} to show that (for given $h$) the fluid current ${\bf j}$ is fully represented by the vorticity and compression fields $\Omega,Q$. To substitute the latter as the fundamental phase space variables, one first changes variables via
\begin{equation}
\prod_l dh_l \, d{\bf j}_l = \prod_l \rho_0^2 h_l^2 dh_l \, d{\bf v}_l.
\label{A23}
\end{equation}
With $\omega = \nabla \times {\bf v} = -\nabla^2 \psi^V$ and $q = \nabla \cdot {\bf v} = -\nabla^2 \phi$, one may further change variables ${\bf v} \to (\omega,q,{\bm \psi}^0)$, in which ${\bm \psi}^0 = \{\psi^0_m \}$ represents the potential flow component ${\bf v}^P = \nabla \times \psi^P$, with an overall constant Jacobian $J_0 = \Delta x^{2N_E}$ (where $N_E = A_D/\Delta x^2$ is the number of spatial grid cells):
\begin{eqnarray}
\prod_l h_l^2 dh_l \, d{\bf v}_l
&=& \prod_{m=2}^{N_D} d\psi^0_m \,
\prod_l \Delta x^2 h_l^2 dh_l \, dq_l \, d\omega_l
\nonumber \\
&=& \prod_{m=2}^{N_D} d\psi^0_m \,
\prod_l \Delta x^2 h_l^4 dh_l \, dQ_l \, d\Omega_l,
\nonumber \\
\label{A24}
\end{eqnarray}
where in the last expression we have introduced $\Omega_l = (\omega_l + f_l)/h_l$, $Q_l = q_l/h_l$. One obtains finally:
\begin{eqnarray}
d\Gamma_E &=& \rho({\cal H},\{{\cal C}_i\})
\prod_{m=2}^{N_D} \frac{d\psi^0_m}{P_0 H_0}
\nonumber \\
&&\times\ \lim_{\Delta x \to 0} \prod_l \frac{\rho_0^2 \Delta x^2}{H_0 P_0^2}
h_l^4 dh_l \, dQ_l \, d\Omega_l.
\label{A25}
\end{eqnarray}
This is the basic form that is used in Sec.\ \ref{sec:statmech} to derive the shallow water equilibrium equations.

\section{Liouville theorem and inequivalent phase space measures}
\label{app:liouvilleinequiv}

Given its centrality to the differences between the present work and that of RVB \cite{RVB2016}, for completeness we carefully summarize here, at a more general level, the constraints enforced by the Liouville theorem on the form of the infinite-dimensional invariant phase space measure. It is shown that the continuum limit obtained from the fluid parcel \emph{area} discretization consistent with the theorem proven in App.\ \ref{app:liouville} is explicitly inequivalent to the fluid parcel \emph{volume} discretization adopted by RVB.

To focus the discussion, consider a phase space $\Gamma$ defined by a single continuous scalar field ${\bm \varphi} \equiv \{\varphi({\bf r}) \}_{{\bf r} \in D}$, with $D$ a spatial domain (2D in this case). The generalization to multiple fields simply adds more indices. The field is assumed to obey a first order equation of motion of the form
\begin{equation}
\partial_t \varphi({\bf r}) = V({\bf r};{\bm \varphi}),
\label{B1}
\end{equation}
where ${\bf V}[{\bm \varphi}] = \{V({\bf r};{\bm \varphi})\}_{{\bf r} \in D}$ is the phase space flow field, each of whose ``components'' ${\bf r}$ is a functional of ${\bf \varphi}$. A probability density $P[{\bm \varphi}]$ on the phase space obeys an equation of motion
\begin{equation}
\partial_t P + \nabla_{\bm \varphi} \cdot (P {\bf V}) = 0,
\label{B2}
\end{equation}
in which the explicit form of the phase space divergence of any vector ${\bf F}$ is defined by the functional derivative
\begin{equation}
\nabla_{\bm \varphi} \cdot {\bf F}[{\bm \varphi}] = \int_D d{\bf r}
\frac{\delta F({\bf r};{\bm \varphi})}{\delta \varphi({\bf r})}.
\label{B3}
\end{equation}
Conservation of probability follows by integrating (\ref{B2}) over $\Gamma$ and applying the infinite dimensional Gauss law to the second term:
\begin{equation}
\partial_t \int_\Gamma {\cal D}[{\bm \varphi}] P[{\bm \varphi}]
= -\int_{\partial \Gamma} P[{\bm \varphi}] {\bf V}[{\bm \varphi}]
\cdot d{\bm \Sigma}[{\bm \varphi}] \to 0,
\label{B4}
\end{equation}
where $d{\bm \Sigma}$ is the outward pointing area element. The vanishing of the right hand side is based on the assumption that $P$ vanishes as the boundary $\partial \Gamma$ is pushed to infinity (or, in some models with bounded fields, that there is a well defined finite boundary through which the flows do not pass, ${\bf V}[{\bm \varphi}] \cdot d\Sigma[{\bm \varphi}] = 0)$.

Consistent with the form of (\ref{B3}), and the subsequent application of Gauss's law, the phase space functional integral underlying averages with respect to $P$ must be defined by free integration over each field component:
\begin{equation}
\int {\cal D}[{\bm \varphi}] = \prod_{\bf r} \int d\varphi({\bf r})
\label{B5}
\end{equation}
The right hand side is most conveniently defined by first approximating $D$ by a regular, uniform grid $\{{\bf r}_i\}_{i=1}^N$ of mesh size $b$ and then taking the continuum limit,
\begin{equation}
\int {\cal D}[{\bm \varphi}] = \lim_{b \to 0}
\frac{1}{{\cal N}(b)} \prod_{i=1}^N \int d\varphi_i,
\label{B6}
\end{equation}
where ${\cal N}(b)$ is an overall normalization. Of course, other, non-uniform grids may be chosen, but they are constrained by the requirement that they give the same continuum limit, e.g., for physically well defined statistical averages. Rigorous examples of this construction include Feynman path integrals, and higher dimensional random surface integrals, that may defined through Brownian motion Wiener measures. It should be emphasized here that there is nothing in principle that forbids a grid choice that gives a physically different continuum limit from generating mathematically consistent probabilities, but it must correspond to a different dynamical model than (\ref{B1}).

This association between the form of the divergence and the integration measure is critical to proper application of the Liouville theorem. Specifically, the Liouville theorem holds if the phase space flows can be shown to be divergence-free:
\begin{equation}
\nabla_{\bm \varphi} \cdot {\bf V}[{\bm \varphi}] = 0.
\label{B7}
\end{equation}
This follows trivially for Hamiltonian flows, where ${\bm \varphi} = ({\bf q},{\bf p})$ is composed of conjugate pairs of coordinate and momentum variables. Then
\begin{equation}
{\bf V}({\bf q},{\bf p}) = (\nabla_{\bf p} H, -\nabla_{\bf q} H),
\label{B8}
\end{equation}
is derived from a Hamiltonian $H({\bf q},{\bf p})$, and
\begin{equation}
\nabla_{\bm \varphi} \cdot {\bf V}
= \nabla_{\bf q} \cdot \nabla_{\bf p} H
- \nabla_{\bf p} \cdot \nabla_{\bf q} H = 0.
\label{B9}
\end{equation}

When (\ref{B7}) is satisfied, it follows from (\ref{B3}) that the probability density is freely advected by the flow,
\begin{equation}
\partial_t P + {\bf V} \cdot \nabla_{\bm \varphi} P = 0,
\label{B10}
\end{equation}
and in particular, an invariant measure $P_0$, for which $\partial_t P_0 = 0$, must obey \begin{equation}
{\bf V} \cdot \nabla_{\bm \varphi} P_0[{\bm \varphi}] = 0.
\label{B11}
\end{equation}
However, one then notes that
\begin{equation}
\frac{d}{dt} P_0[{\bm \varphi}]
= \nabla_{\bm \varphi} P_0[{\bm \varphi}] \cdot \frac{d}{dt} {\bm \varphi}
= {\bf V}[{\bm \varphi}] \cdot \nabla_{\bm \varphi} P_0[{\bm \varphi}] = 0,
\label{B12}
\end{equation}
showing that $P_0[{\bm \varphi}]$ is conserved by the flow, and hence one may write $P_0[{\bm \varphi}] = \rho(\{C_l[{\bm \varphi}]\})$, an arbitrary (normalized) function of the set of all integrals of the motion $\{C_l\}$. The microcanonical ensemble adopted in \cite{RVB2016} corresponds to the choice
\begin{equation}
\rho = \prod_l \delta(c_l - C_l[{\bm \varphi}]).
\label{B13}
\end{equation}

One may compare the above formulation to one in which, instead, the Liouville theorem takes the form
\begin{equation}
\int_D d{\bf r} w({\bf r})
\frac{\delta V({\bf r};{\bm \varphi})}{\delta \varphi({\bf r})} = 0,
\label{B14}
\end{equation}
where $w$ is a positive function that is in general a functional of ${\bm \varphi}$---it corresponds to the fluid height field $h$ in the shallow water application. One may construct a domain (but not area) preserving map ${\bf r}({\bf a}): D \to D$ with Jacobian
\begin{equation}
J({\bf a}) \equiv \left|\frac{\partial {\bf r}}{\partial {\bf a}} \right|
= \frac{w_0}{w[{\bf r}({\bf a})]},\ \
w_0 \equiv \frac{1}{A_D} \int_D d{\bf r} w({\bf r}),
\label{B15}
\end{equation}
where $A_D$ is the area of $D$, and with corresponding fields $\tilde \varphi({\bf a}) \equiv \varphi[{\bf r}({\bf a})]$ and $\tilde {\bf V}({\bf a};\tilde {\bm \varphi}) = {\bf V}[{\bf r}({\bf a}); {\bm \varphi}|_{{\bf r} \to {\bf r}({\bf a})}]$. One then obtains
\begin{equation}
\int_D d{\bf r} w({\bf r})
\frac{\delta V({\bf r};{\bm \varphi})}{\delta \varphi({\bf r})}
= \int_D d{\bf a}
\frac{\delta \tilde V({\bf a};\tilde {\bm \varphi})}
{\delta \tilde \varphi({\bf a})} = 0,
\label{B16}
\end{equation}
corresponding to a standard Liouville theorem in the new variables. It follows, according to (\ref{B5}) and (\ref{B6}), that the continuum limit should be obtained using a uniform grid in the ${\bf a}$ coordinate, hence to a nonuniform grid in the ${\bf r}$ coordinate, with elements of area $b^2 w_0/w[{\bf r}({\bf a}_i)]$. In the context of the shallow water equations, this corresponds precisely, as stated, to the equal fluid volume element choice.

Because the height field fluctuates so strongly in the shallow water model, this constitutes a huge effect that demonstrably changes the equilibrium flow equations. As an illustrative example, consider a Gaussian free energy of the form
\begin{eqnarray}
F(\beta) &=& -\ln\left[\int {\cal D}[{\bm \varphi}]
e^{-\beta H[{\bm \varphi}]} \right]
\nonumber \\
H[{\bm \varphi}] &\equiv& \frac{1}{2} \int_D
d{\bf r} w({\bf r}) \varphi({\bf r})^2
\label{B17}
\end{eqnarray}
The equal area discretization produces (up to a normalization constant)
\begin{equation}
F(\beta) = \frac{1}{2} \int_D d{\bf r} \ln[\beta w({\bf r})/2\pi]
\label{B18}
\end{equation}
while the equal volume discretization produces the manifestly different result
\begin{equation}
F(\beta) = \frac{1}{2} A_D \ln[\beta w_0/2\pi].
\label{B19}
\end{equation}

\section{Kinetic energy and momenta in terms of basic fields}
\label{app:KEPi}

Although not needed for the statistical mechanics treatment, for completeness we provide here explicit expressions for the kinetic energy (\ref{3.21}) and kinetic part of the momentum (\ref{3.22}) in terms of the basic fields $\Omega,Q,h$.

Substituting the velocity decomposition (\ref{3.8}), along with the expressions (\ref{3.10}) and (\ref{3.11}) for the potentials, into (\ref{3.21}) one obtains for the kinetic energy
\begin{eqnarray}
E_K = E_K^{VC} + E_K^{VCP} +  E_K^P,
\label{C1}
\end{eqnarray}
where the vortical and compressional components are encompassed by the term
\begin{eqnarray}
E_K^{VC} &=& \frac{\rho_0}{2} \int_D d{\bf r} h({\bf r})
\left|\nabla \times \psi^V({\bf r}) - \nabla \phi({\bf r}) \right|^2
\label{C2} \\
&=& \frac{\rho_0}{2} \int_D d{\bf r}' \int_D d{\bf r}'
\left[\begin{array}{c}
\omega({\bf r}) \\ q({\bf r})
\end{array} \right]^T
\hat {\cal G}_h({\bf r},{\bf r}')
\left[\begin{array}{c}
\omega({\bf r}') \\ q({\bf r}')
\end{array} \right],
\nonumber
\end{eqnarray}
in which the tensor Green function, which depends on the height field, is defined by
\begin{widetext}
\begin{equation}
\hat {\cal G}_h({\bf r},{\bf r}') = \int_D h(\bar {\bf r}) d\bar {\bf r}
\left[\begin{array}{cc}
\bar \nabla G_D(\bar {\bf r},{\bf r}) \cdot  \bar \nabla G_D(\bar {\bf r},{\bf r}') &
\bar \nabla G_D(\bar {\bf r},{\bf r}) \times \bar \nabla G_N(\bar {\bf r},{\bf r}') \\
-\bar \nabla G_N(\bar {\bf r},{\bf r}) \times \bar \nabla G_D(\bar {\bf r},{\bf r}') &
\bar \nabla G_N(\bar {\bf r},{\bf r}) \cdot  \bar \nabla G_N(\bar {\bf r},{\bf r}')
\end{array} \right].
\label{C3}
\end{equation}
%
\end{widetext}
The potential term is given by
\begin{eqnarray}
E_K^P &=& \frac{\rho_0}{2} \int d{\bf r}
h({\bf r}) |{\bf v}^P({\bf r})|^2
\nonumber \\
&=& \frac{\rho_0}{2} \sum_{l,m=2}^{N_D}
\Gamma_{lm}[h] (\psi_l^0 - \psi_1^0)(\psi_l^0 - \psi_1^0)
\label{C4}
\end{eqnarray}
where
\begin{equation}
\Gamma_{lm}[h] = \Gamma_{ml}[h] = \int d{\bf r} h({\bf r})
{\bf v}^P_l({\bf r}) \cdot {\bf v}^P_m({\bf r})
\label{C5}
\end{equation}
is also a (linear) functional of the height field, and does not simplify to a pure circulation integral. Finally, the cross term may be written in the form
\begin{eqnarray}
E^{VCP}_K &=& \rho_0 \int_D d{\bf r} {\bf v}^P({\bf r})
\cdot [\nabla \times \psi^V({\bf r}) - \nabla \phi({\bf r})]
\nonumber \\
&=& - \rho_0 \sum_{l=2}^{N_D} (\psi_l^0 - \psi_1^0)
\nonumber \\
&&\times\ \int d{\bf r} [H^V_l({\bf r}) \omega({\bf r})
+ H^C_l({\bf r}) q({\bf r})],
\label{C6}
\end{eqnarray}
where
\begin{eqnarray}
H^V_l({\bf r}) &=& -\int_D h(\bar {\bf r}) d\bar {\bf r}
{\bf v}^P_l(\bar {\bf r}) \times \bar \nabla G_D(\bar {\bf r},{\bf r})
\nonumber \\
H^C_l({\bf r}) &=& \int_D h(\bar {\bf r}) d\bar {\bf r}
{\bf v}^P_l(\bar {\bf r}) \cdot \bar \nabla G_N(\bar {\bf r},{\bf r})
\label{C7}
\end{eqnarray}
are inner products between the potential velocity field eigenfunction and velocity fields generated by either unit point vortex or unit compression at ${\bf r}$. Both are linear functionals of the height field, and therefore, unlike (\ref{3.9}), do not vanish.

Substituting the velocity decomposition into (\ref{3.22}), one obtains
\begin{eqnarray}
\Pi_K &=& \Pi_K^{VC} + \Pi_K^P
\nonumber \\
\Pi_k^{VC} &=& -\rho_0 \int_D d{\bf r}
\left[H^V_\Pi({\bf r}) \omega({\bf r})
+ H^C_\Pi({\bf r}) q({\bf r}) \right]
\nonumber \\
\Pi_K^P &=& \rho_0 \sum_{l=2}^{N_D}
\Gamma^\Pi_l (\psi_l^0 - \psi_1^0)
\label{C8}
\end{eqnarray}
where
\begin{eqnarray}
H^V_\Pi({\bf r}) &=& -\int_D h(\bar {\bf r}) d\bar {\bf r}
{\bf v}_\Pi(\bar {\bf r}) \times \nabla G_D(\bar {\bf r},{\bf r})
\nonumber \\
H^C_\Pi({\bf r}) &=& \int_D h(\bar {\bf r}) d\bar {\bf r}
{\bf v}_\Pi(\bar {\bf r}) \cdot \nabla G_N(\bar {\bf r},{\bf r})
\nonumber \\
\Gamma^\Pi_l &=& \int_D h({\bf r}) d{\bf r}
{\bf v}_\Pi({\bf r}) \cdot {\bf v}^P_l({\bf r}).
\label{C9}
\end{eqnarray}
Note that all of these are also linear functionals of the fluctuating height field $h$.

Substituting $\omega = h\Omega+f$, $q = hQ$ provides the desired representation in terms of $\Omega,Q,h$.


\begin{thebibliography}{}
\bibitem{foot:2Dcompress} An alternative, entirely equivalent, interpretation of the shallow water equations is that of a purely 2D fluid with density $h$, compressibility $g$, and background potential $h_b$. Surface waves are now acoustic compressional waves.
\bibitem{foot:Qint} Note that the result equivalent to $\Phi \equiv 0$ could have been obtained in principle by integrating $Q$ out directly in (\ref{4.4}), and then performing a KHS transformation on $\Omega$ alone. However, it is useful to exhibit the symmetric appearance of $\Omega$ and $Q$, with the major difference being only that there is no Lagrange multiplier function analogous to $\mu$ controlling the behavior of $Q$. The above model containing $\Omega,h$ alone, is derived in Sec.\ \ref{subsec:nonlinmembrane} via the reverse procedure, undoing the KHS transformation.
\bibitem{foot:C} Recall here that $\Psi^\mathrm{eq}$ is real, but the contour $C$ dictates that $\delta \Psi$ be pure imaginary, ensuring that $e^{\bar \beta W} \sim e^{\bar \beta |\nabla (i \delta \Psi)|^2/2\rho_0 h}$ is rapidly decreasing for large $|\delta \Psi|$. Note that the $h$-integral contained in the definition of $W$ converges only if the real part of $\xi = -|\nabla \Psi|^2$ is positive, and this is guaranteed for all $\delta \Psi$ only if $C$ coincides with the imaginary axis. The shifted coutour therefore, as is standard in steepest descent calculations, relies on analytic continuation of $W$ to $\mathrm{Re} \, \xi < 0$. As an example, the result $\int_0^\infty d\lambda e^{-(\xi/\lambda + \lambda)} = \sqrt{\xi} K_1(\sqrt{\xi})$ extends the definition of the integral to any complex value of $\xi$ \cite{GR1980}.
\bibitem{foot:QRVB} Alternatively, RVB replace $\bar Q$ by the nonlocal quantity $\mu = (-\nabla^2)^{-1/2} q = (-\nabla^2)^{1/2} \phi$ which also produces a $\Delta x$ scaling. Their formulation also produces a well defined Gaussian distribution in the continuum limit, which is also statistically independent of the other fields. In the present formulation one could introduce the analogous form $\mu = (-\nabla \cdot h \nabla)^{1/2} \phi$, which may be useful for some purposes [e.g., formally integrating out $Q$ in (\ref{4.4}) without using the KHS transformation], but does not produce statistical independence.
\bibitem{foot:liouville} The functional derivative in (\ref{6.8}) is quite singular, including divergent terms $\delta(0)$ and $\nabla \times G_h({\bf r},{\bf r}')|_{{\bf r}'={\bf r}}$. However, this form may be regularized to produce the desired limiting result. The underlying physical reasoning is that the self-advection of a vortex, which is measured by the (\ref{6.8}), does not come from the singular vortex core, but only from ``image flows'' generated by the boundary conditions and by the bottom topography. This allows one, for example, to subtract off the logarithmic singularity in $G_h$.
\end{thebibliography}
\end{document}